\documentclass{article}
\pdfoutput=1 

\usepackage{jcappub} 
\usepackage{aasmacros}

\usepackage{graphicx, epsfig,dsfont,subfigure}
\usepackage{multirow,color} 

\def\case#1/#2{\frac{#1}{#2}}

\def \data {}
\newcommand{\be}{\begin{equation}}
\newcommand{\ee}{\end{equation}}
\newcommand{\bea}{\begin{eqnarray}}
\newcommand{\eea}{\end{eqnarray}}

\def\rf#1{(\ref{#1})}

\def\de#1/de#2{\frac{\partial {#1}}{\partial {#2}}}

\title{\boldmath Non minimally coupled condensate cosmologies: matching observational data with phase space}

\author[a]{S.Carloni}
\affiliation[a]{Centro Multidisciplinar de Astrofisica - CENTRA,
Instituto Superior Tecnico - IST,\\
Universidade de Lisboa - UL,
Avenida Rovisco Pais 1, 1049-001, Lisbon (Portugal)}
\emailAdd{csante.carloni@tecnico.ulisboa.pt}
\author[b]{R. Cianci}
\affiliation[b]{DIME Sez. Metodi e Modelli Matematici, Universit\`{a} di Genova, Via All'Opera Pia 15, 16145 - Genoa, (Italy).}
\emailAdd{cianci@dime.unige.it}
\author[b,d]{P. Feola}
\emailAdd{feola@dime.unige.it}
\author[c,d]{E. Piedipalumbo}
\affiliation[c]{Dipartimento di Fisica, Universit\`{a} degli Studi di Napoli Federico II, Compl. Univ. Monte S. Angelo, 80126 Naples, Italy}
\affiliation[d]{I.N.F.N., Sez. di Napoli, Compl. Univ. Monte S. Angelo, Edificio 6, via Cinthia, 80126 - Napoli, Italy.}
\emailAdd{ester@na.infn.it}
\note{author for correspondence: E.Piedipalumbo}
\author[b]{S.Vignolo}
\emailAdd{vignolo@dime.unige.it}

\abstract{We compare theoretical predictions with observations for a class of cosmological models in which the  dark energy component  is modeled as a fermionic condensate, non-minimally coupled with the gravitational field and characterized by some specific self-interaction potentials. Our analysis is based on  the Markov Chain Monte Carlo Method (MCMC) and employs different data sets.  It turns out that  with an appropriate choice of parameters our models are fully compatible with several observed data. We combine these parameter values with phase space analysis to deduce the features of the entire cosmic history of the considered models.}
\begin{document}

 \maketitle

\section{Introduction}
In the last  years, a large quantity of observational data revealed that the present universe is experiencing an accelerated expansion (see for instance \cite{Riess07, Union2.1,PlanckXXVI, PlanckXIII}), which is  usually assumed to be driven by an new form of matter--energy: dark energy. This latter is in general considered as a cosmic fluid with negative pressure which, according to the most recent estimates, should provide  about $70\% $ of the total amount of the matter-energy in the universe, so representing today the dominant component. 

The nature of dark energy is still completely unknown. Some of the theoretical frameworks proposed to understand this issue include a non zero cosmological constant, the potential energy of some scalar field, effects connected with non homogeneous distributions of matter and averaging procedures, and effects due to alternative theories of gravity. Among the latter, the so called scalar-tensor gravitational theories, arising in other contexts like for example the low energy limit of Kaluza-Klein gravity or the quantum field theory in curved spacetimes \cite{Overduin,Birrell}, have been widely investigated in both the scenarios of the early and late universe expansions. The peculiar feature of these theories is the non minimal coupling (NMC) of gravity with a given scalar field. Recently, it is has been suggested that such a scalar field is not fundamental but it may be constituted by a fermion condensate. 

In cosmology, fermion fields  have been studied as possible sources of inflation and dark energy \cite{Picon,Saha,Saha1,Ribas1}. In most cases, fermions fields are minimally coupled to gravity; only recently, few works have instead investigated the cosmological effects of non--minimally coupled condensates of semi--classical fermions \cite{Souza,Ribas,Grams,CVC,VCF}. Differently from other scalar-tensor models, the fermion condensate are characterized by an easier first order evolution equation. This induces a number of peculiar features which have been described for the first time via phase space analysis in \cite{CVC} for a non--minimally coupled fermion condensate characterized by three different potentials. The dynamical system approaches and methods that have been used in \cite{CVC} have been known for long time in GR based cosmologies \cite{ellisbook}, and more recently in the context of modified gravity (see \cite{Bahamonde:2017ize} for a recent review), allowing a relatively easy semi-quantitative interpretation of complex cosmological models. 

In this paper we aim to perform a comparison of the models introduced in \cite{CVC} with observations and more specifically with the Union2 Type Ia Supernovae (SNIa) data set, the GRBs Hubble diagram, a sample of 28  measurements of the Hubble parameter compiled in \cite{farooqb}, the gaussian priors on the distance from the Baryon Acoustic Oscillations (BAO), and the Hubble constant $h$ (such priors have been included in order to help break the degeneracies among model parameters). The statistical analysis on these datasets is based on Monte Carlo Markov Chains (MCMC) simulations which allow us to compute, simultaneously, the full probability density functions (PDFs) of all the parameters of interest. We show that a value of these parameters exists which is able to match these data to the non-minimally coupled condensate theory. Their values are then used in combination with phase space analysis to deduce global features of the cosmic history suggested by these theories.  For this purpose, we give here a set of dynamical system variables different from the ones employed in \cite{CVC}. Such new setting is required because the  the phase space description in \cite{CVC} was not compact and the data analysis suggests that the present universe is on an orbit that would go to the infinite boundary. The new formulation, instead, is compact and allows us to explore without problems the entire cosmic history selected by the values the observational parameters.

The paper is organized as follow. In Section II, we briefly review the general properties of the theory considered in this paper and its key equations in the Friedmann-Lemaitre-Robertson-Walker (FLRW) metric. In section III, we derive the system of cosmological equations in terms of the redshift {\it z} so as to plot the Hubble's function in terms of the redshift. In section IV, we consider specific cosmological models associated with three different self-interaction potentials: a power law, an exponential and an extended power law potential. In section V, we use observational data sets on SNIa and GRB Hubble diagram, setting Gaussian priors on the distance data from the BAO and the Hubble constant. In section VI, we perform a statistical analysis of the cosmological parameters, based on Monte Carlo Markov Chains (MCMC) simulations.  In section VII, we compare of our fermionic models with the Chevallier-Polarski-Linder (CPL) model. In Section VIII, we combine the above data analysis with the dynamical system approach. Finally, we devote Section IX to conclusions. 

Unless other specified, Latin and Greek indices run from 0 to 3; $\nabla$ denotes the Levi-Civita covariant derivative associated with a metric tensor $g_{ij}$ of signature $(+,-,-,-)$; the Riemann tensor is defined by
\begin{equation}\label{Riemann}
R^d_{cab} = \partial_a\Gamma^{\;\;\;d}_{bc}-\partial_b\Gamma^{\;\;\;d}_{ac}+\Gamma^{\;\;\;d}_{ap}\Gamma^{\;\;\;p}_{bc}-\Gamma^{\;\;\;h}_{bp}\Gamma^{\;\;\;p}_{ac}, 
\end{equation}
where the $\Gamma^{\;\;\;c}_{ab}$ are the Christoffel symbols associated with the metric $g_{ij}$; the Ricci tensor is obtained by contracting the first and the third index via the metric $g_{ab}$:
\begin{equation}\label{Ricci}
R_{ab}=R^c_{acb}.
\end{equation}
\section{The $(1+\epsilon(\bar\psi\psi))R$-theory in a FLRW metric}
For convenience of the reader, we briefly review the main features of the theory proposed in \cite{CVC}. Making use of natural units $(\hslash=c=k_{B}=8\pi G=1)$ for simplicity, let us consider the action functional
\begin{equation}\label{1.1}
{\cal S}= \int{\sqrt{|g|}\left[\left(1 + \epsilon(\bar\psi\psi)\right)\/R - L_D\right]ds},
\end{equation}
in which the Einstein--Hilbert term is non--minimally coupled to the condensate $\bar\psi\psi$ associated with a Dirac field. The latter has a Lagrangian of the usual form
\be\label{2.1}
L_D = \frac{i}{2}\left( \bar{\psi}\Gamma^iD_i\psi - D_i\bar{\psi}\Gamma^i\psi\right) -m\bar{\psi}\psi + V(\bar\psi\psi),
\ee
where a fermionic self--interaction potential $V(\bar\psi\psi)\/$ is present. In eq.~\eqref{1.1} $\epsilon$ indicates a suitable constant parameter, while in eq.~\eqref{2.1} we have $\Gamma^i :=e^i_\mu\gamma^\mu\/$, $\gamma^\mu\/$ representing Dirac matrices and $e^\mu_i\/$ a tetrad field such that the metric tensor can be expressed as $g_{ij}=e^\mu_ie^\nu_j\eta_{\mu\nu}$ ($\eta_{\mu\nu}={\rm diag.}(1,-1,-1,-1)$); $D_i\/$ denotes the covariant derivative of the spinor field 
\begin{equation}\label{3.1}
	D_i\psi = \partial_i\psi - \Omega_i\psi, \qquad
	D_i\bar{\psi} = \partial_i\bar{\psi} + \bar{\psi}\Omega_i,
\end{equation}
where
\begin{equation}\label{4.1}
\Omega_i = -\frac{1}{4}g_{ij}\left(\Gamma^{\;\;\;j}_{pq} - e^j_\mu\partial_pe^\mu_q\right)\Gamma^p\Gamma^q
\end{equation}
are the spin covariant derivative coefficients associated with the Levi--Civita connection $\Gamma^{\;\;\;j}_{pq}$.

 Some remarks are here in order on the nature of the fermion field we are considering. Our approach is entirely classical, $\psi$ denoting a set of four complex--valued spacetime functions which transform according to the spinor representation of the Lorentz group. This means that, rather than an actual fermion field, $\psi$ can be though to represent macroscopic objects made of fermions (see e.g. \cite{Picon}).  Indeed, as we are considering cosmic eras in which particles are non-relativistic, a second quantization fermion field would not really be compatible with our assumptions. On top of this, we  are restricting specifically to the case in which such fermion field is a condensate and it is known that in this case  a classical description of the field  is able to capture all the key properties of the quantum field \cite{Picon}.

As a consequence, our Dirac action and the resulting matter field equations do not necessarily have to satisfy the same properties as those in quantum field theory.  This applies, for example to the issue of renormalizability. As a matter of fact, if we were to insist considering $\psi$ a quantum field, we could ask if the presence of the self-interaction potential and even more the non-minimal coupling can make the Dirac equations non--renormalizable.  In this regards, in Section \ref{models} we shell  consider three different potentials (see subsequent Eqs. (\ref{pot1}), (\ref{pot2}), (\ref{pot3})). In particular,  in the minimal coupled theory the potentials (\ref{pot1}) and (\ref{pot3}) would be renormalizable respectively for $\alpha < \frac{4}{3}$ and $\gamma < \frac{2}{3}$, while potential (\ref{pot2}) would be always renormalizable.Instead, the real problem lays, at least in the purely metric theory, in the non--renormalizable terms generated by  the non-minimal coupling. A possible way out of this problem would be to introduce torsion. Indeed, following the procedure given in \cite{FVC} and according to Wilson’s analysis of renormalizability, it is easy to see that in the theory with torsion the presence of non-minimal coupling leads always to renormalizable actions, provided that the considered potentials have the values of $\alpha$ and $\gamma$ in the intervals mentioned above. We will see that the values we will obtain for these parameters belong to these intervals.
 
Making use of the notation $\varphi := \bar\psi\psi$, from the the action \eqref{1.1} we derive Einstein--like and Dirac equations respectively of the form
\begin{equation}\label{5.1}
\left(1 + \epsilon\varphi\right)\left(R_{ij} - \frac{1}{2}Rg_{ij}\right)= \Sigma_{ij} + \epsilon\left(\nabla_i\nabla_j\/\varphi - g_{ij}g^{pq}\nabla_p\nabla_q\/\varphi\right),
\end{equation}
and
\begin{subequations}\label{7.1}
\begin{equation}\label{7.1a}
 i\Gamma^iD_i\psi -m\psi + V'(\varphi)\psi - \epsilon\psi\/R =0,
 \end{equation}
 \begin{equation}\label{7.1b}
 iD_i\bar{\psi}\Gamma^i + m\bar\psi - V'(\varphi)\bar\psi + \epsilon\bar\psi\/R =0,
\end{equation}
\end{subequations}
where
\be\label{6.1}
\Sigma_{ij} = \frac{i}{4}\left( \bar{\psi}\Gamma_{(i}D_{j)}\psi - D_{(i}\bar{\psi}\Gamma_{j)}\psi\right) - \frac{1}{2}L_D\,g_{ij},
\ee
is the energy--momentum tensor of the Dirac field and $V' := \frac{dV}{d\varphi}$. For later use, inserting eqs. \eqref{7.1} into \eqref{6.1}, we can express the energy--momentum tensor $\Sigma_{ij}$ as
\be\label{8.1}
\Sigma_{ij} = + \frac{i}{4}\left( \bar{\psi}\Gamma_{(i}D_{j)}\psi - D_{(i}\bar{\psi}\Gamma_{j)}\psi\right) - \frac{\epsilon}{2}\varphi\/R\,g_{ij} -\frac{1}{2}V(\varphi)\,g_{ij} + \frac{1}{2}\varphi\/V'(\varphi)\,g_{ij}.
\ee 
In order to discuss cosmological models arising from the above presented theory, let us consider a spatially flat Friedmann--Lema\^{\i}tre--Robertson--Walker (FLRW) metric tensor
\begin{equation}\label{9.1}
ds^2 = dt^2 - a(t)^2\/\left( dx^2 + dy^2 + dz^2 \right).
\end{equation}
It is seen that the Einstein--like equations \eqref{5.1}, evaluated in the metric \eqref{9.1}, reduce to \cite{CVC}  
\begin{subequations}\label{16.1}
\begin{equation}\label{16.1a}
(1+\epsilon\varphi)\/3\left(\frac{\dot a}{a}\right)^2 = \frac{m}{2}\varphi -3\epsilon\frac{\dot a}{a}{\dot\varphi} - \frac{1}{2}V(\varphi),
\end{equation}
\be\label{16.1b}
(1+\epsilon\varphi)\left[2\frac{\ddot a}{a} + \left( \frac{\dot a}{a} \right)^2\right] = -\frac{\epsilon}{2}\varphi\/R - \epsilon\ddot\varphi - 2\epsilon\frac{\dot a}{a}\dot\varphi -\frac{1}{2}V(\varphi) +\frac{1}{2}\varphi\/V'(\varphi).
\ee
\end{subequations}
We can replace eq.~\eqref{16.1b} by the equivalent Raychaudhuri equation
\be\label{17.1}
(1+\epsilon\varphi)6\frac{\ddot a}{a} = -\frac{3}{2}\epsilon\varphi\/R - 3\epsilon\ddot\varphi - 3\epsilon\frac{\dot a}{a}\dot\varphi - \frac{m}{2}\varphi - V(\varphi) + \frac{3}{2}\varphi\/V'(\varphi).
\ee
Analogously, in the metric \eqref{9.1} the Dirac equations \eqref{7.1} assume the expression 
\begin{subequations}\label{13.1}
\begin{equation}\label{13.1a}
	\dot\psi + \frac{3}{2}\frac{\dot a}{a}\psi + im\gamma^0\psi - V'(\varphi)\gamma^0\psi + i\epsilon\/R\gamma^0\psi =0,
	\end{equation}
	\begin{equation}\label{13.1b}
	\dot{\bar\psi} + \frac{3}{2}\frac{\dot a}{a}\bar\psi - im\bar{\psi}\gamma^0 + V'(\varphi)\bar\psi\gamma^0 - i\epsilon\/R\bar\psi\gamma^0 =0.
\end{equation}
\end{subequations}
From eqs. \eqref{13.1} we immediately derive the evolution law for the scalar field $\varphi =\bar\psi\psi$
\be\label{14.1}
\dot\varphi + 3\frac{\dot a}{a}\varphi =0,
\ee
yielding the final relation
\be\label{18.1}
\varphi =\frac{\varphi_0}{a^3}.
\ee
Since $\varphi\rightarrow0$ when the scale factor grows, it is expected that the non-minimal coupling contributions tend to disappear at cosmological late time. For sake of completeness, we can add a perfect fluid to our cosmological model. To this end, we suppose a barotropic perfect fluid assigned, with equation of state $p=w\rho$ ($w\in [0,1[$) and standard conservation law
\be\label{1.3}
\dot\rho + 3\frac{\dot a}{a}\left(\rho + p\right)=0.
\ee
In such a circumsatnce, the field equations \eqref{16.1} and \eqref{17.1} become respectively 
\begin{subequations}\label{2.3}
\begin{equation}\label{2.3a}
(1+\epsilon\varphi)\/3\left(\frac{\dot a}{a}\right)^2 = \rho + \frac{m}{2}\varphi - \frac{1}{2}V(\varphi) -3\epsilon\frac{\dot a}{a}{\dot\varphi},
\end{equation}
\be\label{2.3b}
(1+\epsilon\varphi)\left[2\frac{\ddot a}{a} + \left( \frac{\dot a}{a} \right)^2\right] = -p -\frac{\epsilon}{2}\varphi\/R -\frac{1}{2}V(\varphi) +\frac{1}{2}\varphi\/V'(\varphi) - \epsilon\ddot\varphi - 2\epsilon\frac{\dot a}{a}\dot\varphi,
\ee
\end{subequations}
and
\be\label{3.3}
(1+\epsilon\varphi)6\frac{\ddot a}{a} = -(\rho + 3p) -\frac{3}{2}\epsilon\varphi\/R - 3\epsilon\ddot\varphi - 3\epsilon\frac{\dot a}{a}\dot\varphi - \frac{m}{2}\varphi -V(\varphi) + \frac{3}{2}\varphi\/V'(\varphi).
\ee
Inserting the evolution equation \rf{14.1} for the scalar field and the expression of the Ricci scalar in flat FLRW spacetime $R=-6\left(\frac{\ddot a}{a}+\frac{\dot{a}^2}{a^2} \right)$ into eqs. \eqref{2.3} and \eqref{3.3}, we get the final system of equations 
\begin{subequations}\label{FinalEqCosm}
\begin{equation}\label{FriedDSA}
3(1-2\epsilon\varphi)\left(\frac{\dot a}{a}\right)^2 = \rho + \frac{m}{2}\varphi - \frac{1}{2}V(\varphi),
\end{equation}
\begin{equation}
6(1-2\epsilon\varphi)\frac{\ddot a}{a} = -(\rho + 3p) - 18\epsilon\varphi\left(\frac{\dot a}{a}\right)^2 - \frac{m}{2}\varphi -V(\varphi) + \frac{3}{2}\varphi\/V'(\varphi),
\end{equation}
\begin{equation}
\dot\varphi + 3\frac{\dot a}{a}\varphi =0,
\end{equation}
\begin{equation}
\dot\rho + 3\frac{\dot a}{a}\left( 1+ w\right)\rho=0,
\end{equation}
\end{subequations}
which represent the starting point of our subsequent analysis.
\section{Cosmological equations in terms of the redshift}
In this section we rewrite the system of equations \eqref{FinalEqCosm} in terms of redshift {\it z}. To start with, let us take the identities $a=\frac{a_{0}}{1+z}$ and $H=\frac{\dot a}{a}$ into account, so that  the system \eqref{FinalEqCosm} assumes the form

\begin{subequations}\label{FinalEqCosm2}
\begin{equation}\label{FriedDSA}
3(1-2\epsilon\varphi) H^2= \rho + \frac{m}{2}\varphi - \frac{1}{2}V(\varphi),
\end{equation}
\begin{equation}\label{FriedDSA2}
6(1-2\epsilon\varphi) H^2+ 6(1-2\epsilon\varphi) \dot{H} = -(\rho + 3p) - 18\epsilon\varphi H^2- \frac{m}{2}\varphi -V(\varphi) + \frac{3}{2}\varphi\/V'(\varphi),
\end{equation}
\begin{equation}\label{3.1c}
\dot{\varphi} + 3H \varphi =0,
\end{equation}
\begin{equation}\label{3.1d}
\dot\rho + 3H\left( 1+ w\right)\rho=0,
\end{equation}
\end{subequations}
From now on we restrict ourselves to a dust filled Universe with $w=0$ and thus $p = 0$. In this case, the last two equations have exact solutions 
\begin{eqnarray}\label{phirho}
&&\varphi(z)=\frac{\varphi_0}{a^3}=\varphi_0 (1+z)^3\\
&&\rho(z)=\frac{\rho_0}{a^3}=\rho_0 (1+z)^3\,,
\end{eqnarray}
thus implying that the fermion condensate evolves as the dark matter energy density composed of dust fluid. 
The Eqs. (\ref{FriedDSA}), and (\ref{FriedDSA2} ) can be slightly manipulated in order to define the effective pressure and energy density of the fermion field
\begin{eqnarray}\label{pressure_rho}
&&p_{\varphi}= - \frac{\varphi  \left[(1-2 \epsilon  \varphi ) V'(\varphi )-2 \epsilon  \varphi  \left(m+\frac{2 \rho_0
   }{\varphi_0 }\right)\right]+V(\varphi ) (4 \epsilon  \varphi -1)}{2(1-2 \epsilon  \varphi )}\\
   &&\rho_{\varphi}=\frac{m\varphi }{2}-\frac{V(\varphi )}{2}\\ \nonumber\,
\end{eqnarray}
These two expressions allow us to define an effective equation of state $w_{\varphi}$, which drives the  behavior of the model in the dark energy dominated era:
\begin{equation}\label{wphi}
w_\varphi= - \frac{\varphi  \left[\left(\varphi _0 - 2\epsilon\varphi\varphi _0\right) V'(\varphi )-2 \epsilon 
   \varphi  \left(m \varphi _0+2 \rho _0\right)\right]+ \varphi _0 V(\varphi ) (4 \epsilon  \varphi
   -1)}{\varphi _0 (1-2 \epsilon  \varphi )[m \varphi -V(\varphi )]}\,.
\end{equation}
Now we can interpret the fermion condensate as the source of an effective $\Lambda$ {\it  term} $\Lambda_{eff}$, by defining 
$\Lambda_{eff}={\rho_{\varphi}\over {F}}$ , being $F(\varphi )= (1-2\epsilon\varphi)$, and the effective
gravitational constant as $G_{eff}= {1\over {F}}$. With these
definitions, the Eqs. \eqref{FriedDSA} and \eqref{FriedDSA2} can be rewritten in terms of $\rho_{\varphi}$ and $p_{\varphi}$ as
\begin{eqnarray}
  3H^2 &=& G_{eff}\rho_m +\Lambda_{eff}\,,\label{he}\\
  2 \dot{H}+ 3 H^2&=& - G_{eff} p_{\varphi} \,, \label{e4}
\end{eqnarray}
Introducing the standard $\Omega$ parameters by
\[ \Omega_{m}={\rho_{m}\over {3H^{2}}}, \quad
\Omega_{\Lambda_{eff}}={\rho_{\varphi}\over {3H^{2}}} \,,\] we get the relation
\begin{eqnarray}
   \Omega_m+\Omega_{\Lambda_{eff}}=F\,.
\end{eqnarray}
Note that in order to write the \eqref{he} and \eqref{e4} we have defined
\be
G_{eff} = \frac{1}{1-2\epsilon\varphi}=\frac{1}{1-2\epsilon\varphi_0 (1+z)^3}\label{Geffeq}
\ee
and therefore we assumed that the denominator $1-2\epsilon\varphi_0 (1+z)^3$  does not vanish. 
In order to analyze the cosmological solutions, we first formulate the Friedmann equation (\ref{he}) in terms of the red-shift $z$, making use of the well known relations
\begin{eqnarray}
&& 1+z=\frac{a_0}{a},\\
 && \frac{d}{dt}=-(1+z)H(z)\frac{d}{dz}\,
 \end{eqnarray}
where the Hubble function $H(z)$ is here expressed as
\begin{equation}\label{hubble}
H(z)=\sqrt{\frac{m\varphi (z)-V(\varphi (z))+2 \rho _0 (z+1)^3}{6-12 \epsilon  \varphi (z)}}\,.
\end{equation}

Let us  subsitute $\rho_0=3 H_0^2 \Omega _m$, $\varphi(z)=\varphi_0(1+z)^3$, and $\varphi_0= \beta \rho_0=3 \beta  H_0^2 \Omega _m$, thus obtaning the expression
\begin{equation}\label{hzz.1}
H(z)=\sqrt{\frac{3 H_0^2 (z+1)^3 (\beta  m+2) \Omega _m-V(z)}{6\left[1-6 \beta  H_0^2 (z+1)^3 \epsilon 
   \Omega _m\right]}}\,,
\end{equation}
In Eq. (\ref{hzz.1}) the fermion condensate  not only acts as dark energy, through its self-interaction potential, but it can also play the role of a dark matter term.
In the following we will consider models for which $\epsilon$ and $\varphi_0$ have the same sign. This anzats can appear ill chosen as it allows the presence of divergences  in the  effective gravitational constant and in the Hubble term (\ref{Geffeq}) and (\ref{hzz.1}). The phase space analysis in Section (\ref{dynamical}) reveals, however, that these singular states are unstable and therefore never reachable dynamically. In addition, in order to avoid possible problems within the numerical codes used to process observational data, we set for $\epsilon$ and $\varphi_0 $ very small values, as, for instance, in Fig. (\ref{fig1_all}). It is worth noting that the {\it critical} quantity is the product $\epsilon\, \varphi_0$; therefore we set $\epsilon = 10^{-7}$: in this way the evolution of the cosmology will be always far from the singular states.

\begin{figure}
\includegraphics[width=12 cm]{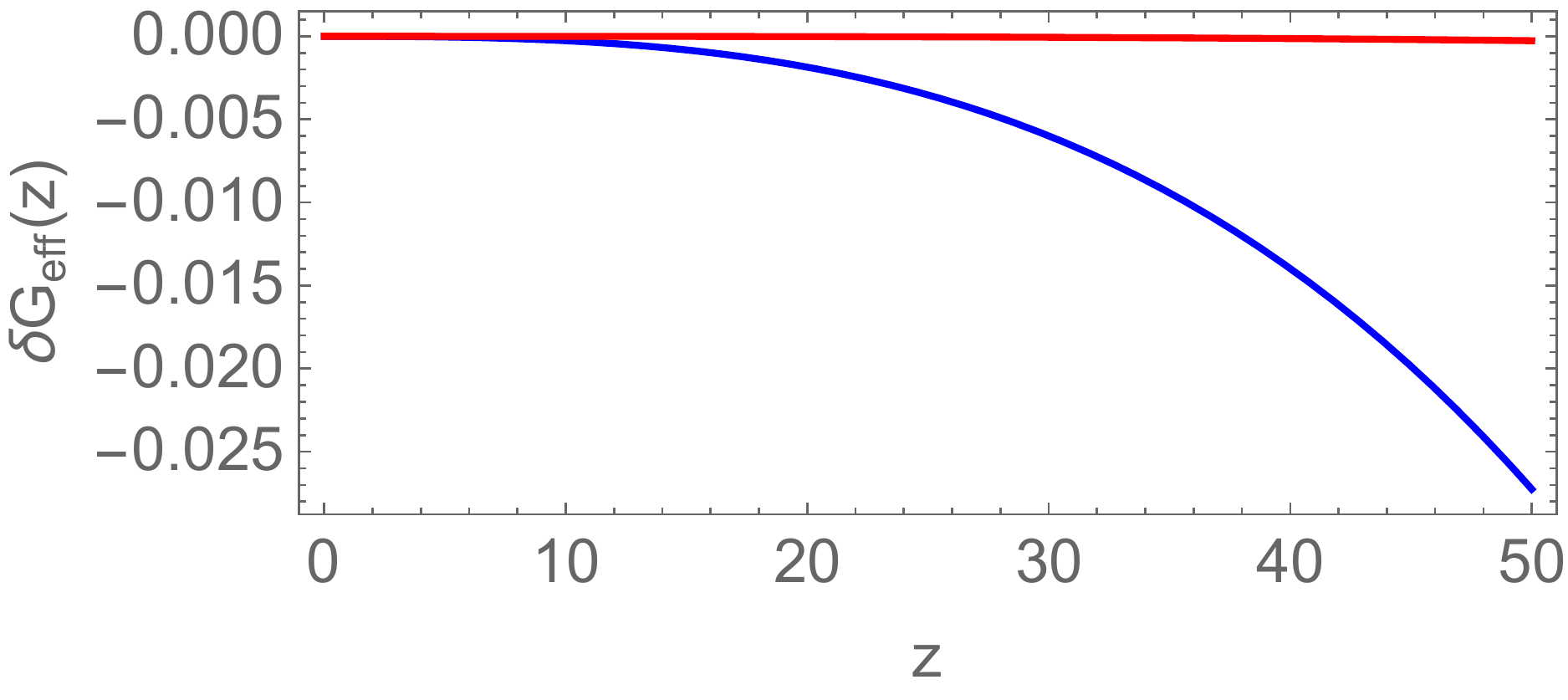}
\caption{{\bf Time evolution for the relative variation $\delta G$ for our model, with $\epsilon=10^{-5}$ (blue line), and $\epsilon=10^{-7}$ (red line). As we see, the evolution of $G_{eff}$ is very different.}}\label{Gvariable1}
\end{figure}

In this regard it is worth illustrating the role of $\epsilon$ on the variation of the effective gravitational constant. Actually, recent data analysis indicates that the gravitational constant and its derivatives are slowly varying,  $i.e. \left|\dot{G}/G\right|\propto 10^{-14}$ \cite{Bambi05, pitjeva05}.  Moreover, a change in $G_{eff}$ would affect observations on solar system dynamics, Big Bang Nucleosynthesis predictions, data concerning the growth rate of structures and the CMB (see e.g. \cite{Bambi05}, \cite{nesseris09}, \cite{ester12}). Our analysis offers a way to probe these differences using measures performed today.
In order to give an idea of the differences that our model could present with respect to the $\Lambda$CDM model, {\bf in Fig. (\ref{Gvariable1}) we plot the behaviour of $\delta G= \frac{G_{eff}-G_N}{G_N}$ for two different values of $\epsilon$. It turns out that the high-z behaviour of the the effective gravitational is different, showing that indeed an importa difference exists between different values of $\epsilon$.}

\section{Three cosmological models}
\label{models}
So far we have made no assumptions about the  form of the self-interaction potential $V(\varphi)$ appearing in the Eqs. \eqref{FriedDSA} and \eqref{FriedDSA2}. In this section we investigate three different forms of $V(\varphi)$:
\begin{eqnarray}
V(\varphi)&=&V_0 \varphi^\alpha \label{pot1}\\
V(\varphi)&=&V_0 \exp(-\lambda \varphi) \label{pot2}\\
V(\varphi)&=&V_0 (\varphi^{2}+V_1)^{\gamma} \label{pot3}\,.
\end{eqnarray}
The  potential $V(\varphi)=V_0 \varphi^\alpha$ is largely used in the dark energy literature and also in the investigation of fermionic dark energy. Because of Eq. \eqref{18.1}, for $a\rightarrow0$ the potential can be, depending on the sign  of the parameter $\alpha$,  negligible or dominant.  The converse happens  when $a\rightarrow\infty$. 
The exponential potentials of the kind \eqref{pot2} are very important to describe not only quintessence models of the dark energy, but also various scenarios of the inflationary expansion of the early universe. Actually, for a flat Friedmann universe filled with a minimally coupled exponential field, we know the general exact solution of the field equations \cite{ACK}. Moreover, in the limit $\varphi\rightarrow0$ exponential potentials become cosmological constant terms, introducing in this way a dynamical realization of the cosmological constant related to the scalar field.
Finally, the potential $V=V_0 (\varphi^{2}+V_1)^\gamma$ is a simple generalization of the power law potential. It has been chosen because of its relevance in inflationary scenarios in the presence of a scalar field. For $a\rightarrow0$ this potential coincides with the pure power law one.  However, for $a\rightarrow\infty$ it generates a cosmological term related to the value of the constant $V_1$. In the following discussion, we will compare theoretical predictions with observational data for each cosmological model arising from the above three different choices of potential. To accomplish this aim, we parametrize in a different way our models introducing in their analytical representation parameters for which we can {\it immagine} the proper ranges of variations\,: for each cosmological model, indeed, $V_0$ is not a fit parameters, but is expressed in terms of different observables, as shown in Eq. (\ref{V0eq}).
\subsection{The case $V(\varphi)=V_0 \varphi^\alpha$}
In order to analyze the cosmological scenarios arising from this power law model, we start from  the Eq. (\ref{hubble}) which can be parametrized in a different way by requiring that $H(0)=H_0$. It turns out that we can express $V_0$ in terms of $H_0$ and the other parameters:
\begin{eqnarray}
\label{V0eq}
V_0=\varphi _0^{-\alpha } \left[6 H_0^2 \left(\Omega _m+2 \epsilon\varphi_0 - 1\right)+m\varphi_0\right]\,,
   \end{eqnarray}
so that  the Hubble function $H$ can be expressed as
\begin{eqnarray}H&=&\sqrt{\frac{(z+1)^3 \left(6 H_0^2 \Omega _m+m \varphi _0\right)-(z+1)^{3 \alpha } \left[6
   H_0^2 \left(\Omega _m+2 \varphi _0 \epsilon -1\right)+m \varphi _0\right]}{6-12
   \varphi _0 (z+1)^3 \epsilon }}  \,.\label{halpha}
\end{eqnarray}
If we introduce the dimensionless Hubble parameter $E(z,{\rm\theta)} =\frac{H(z)}{H_0}$, where ${\rm\theta}$ indicates the set of parameters characterizing the cosmological model, then we can construct the luminosity distance and the modulus of distance according to the following relations:
\begin{eqnarray}
  d_L(z,{\rm\theta}) &=& \frac{c}{H_0} (1+z) \int_0^z \frac{d\zeta}{E(\zeta,{\rm\theta}) }\,,\\ 
  \mu_{th}(z, {\rm\theta})&= &25 + 5 \log{  d_L(z,{\rm\theta}) }\,.
   \end{eqnarray} 
\subsection{The case $V(\varphi)=V_0 \exp(-\lambda \varphi) $}
Following the same procedure implemented in the previous subsection,  also in the case of the exponential potential we use the condition  $H(0)=H_0$ to express $V_0$ in terms of $H_0$ and the other parameters:
\begin{eqnarray}
&&V_0=e^{\lambda  \varphi _0} \left[6
   H_0^2 \left(\Omega _m+2 \varphi
   _0 \epsilon -1\right)+m \varphi
   _0\right]\\ \,,
   &&\end{eqnarray}
so that  the Hubble function $H$ assumes the form
\begin{eqnarray}
&&H(z)=\sqrt{\frac{(z+1)^3 \left(6 H_0^2 \Omega
   _m+m \varphi _0\right)-e^{\lambda
    \varphi _0
   \left[1-(z+1)^3\right]} \left[6
   H_0^2 \left(\Omega _m+2 \varphi
   _0 \epsilon -1\right)+m \varphi
   _0\right]}{6-12 \varphi _0
   (z+1)^3 \epsilon }}
\end{eqnarray}
 \begin{figure}
\includegraphics[width=\textwidth]{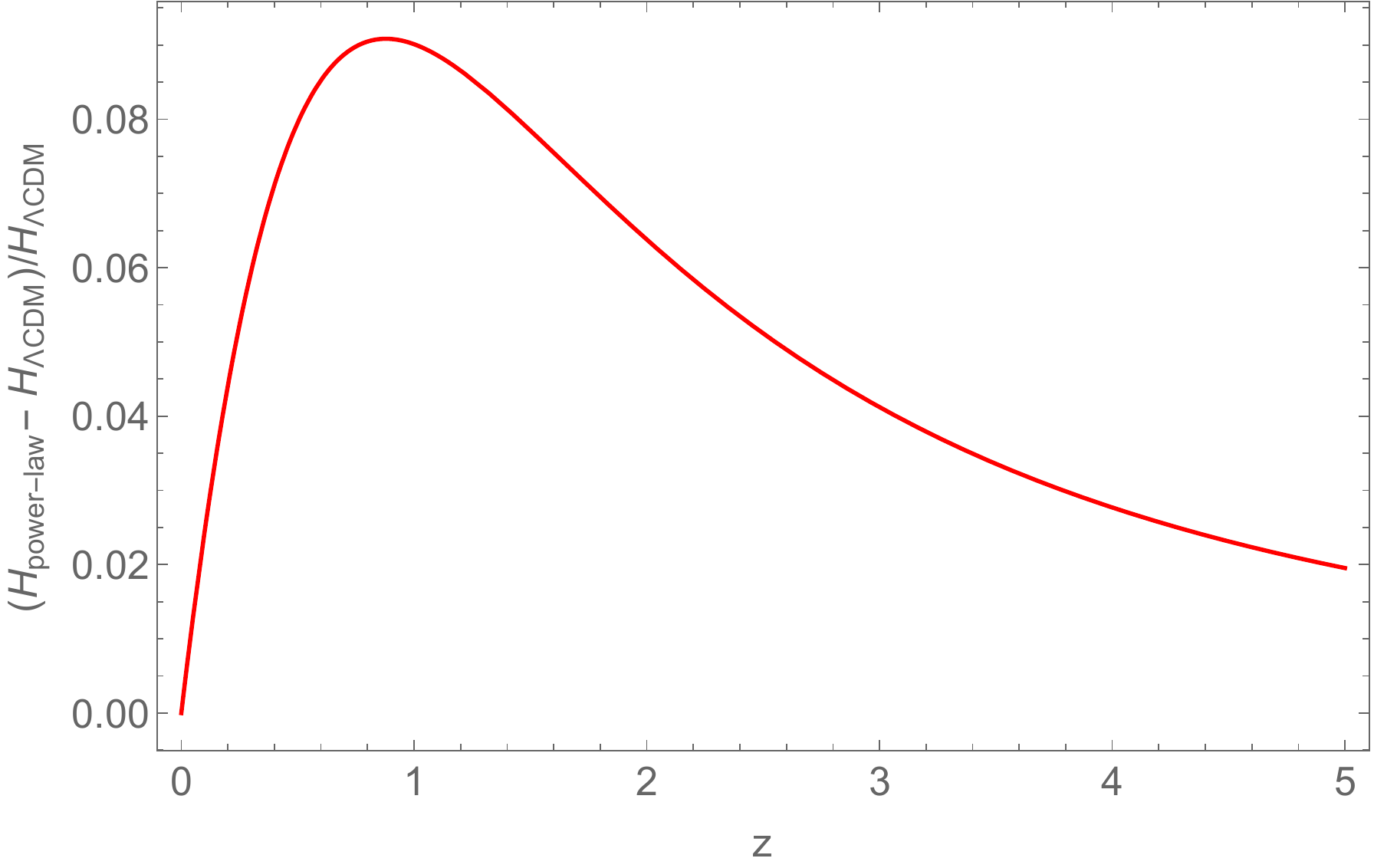}
\caption{ The behaviour in redshift of the relative variation of the Hubble function $H(z)$  for the power law potential  $V(\varphi)=V_0 \varphi^\alpha$, compared with the standard $\Lambda$CDM model. The  parameters correspond to the best fit  values, as in Table (\ref{tabcosmofit}) ($\epsilon=10^{-7}$, $\varphi_0=0.006$, $\alpha=0.23$, $m=1.56$). For this  power law  model we have the worst  matching  with the $\Lambda$CDM.}  
\label{fig1_all}
\end{figure}

\subsection{The case $V(\varphi)=V_0 (\varphi^{2}+V_1)^{\gamma}$}
Owing to the presence of an additional parameter, in this case we include both  $H_0$ and the deceleration parameter $q_0$ in the new parametrization and, consequently, we need more constraints. Actually,  the conditions  $H(0)=H_0$ 
allow us to express  $V_0$ in terms of the other parameters:
\begin{equation}\label{V0eq2}
V_0=\left(V_1+\varphi _0^2\right){}^{-\gamma } \left[6 H_0^2 \left(\Omega _m+2 \epsilon  \varphi
   _0-1\right)+m \varphi _0\right]\,.
\end{equation}
With this substitution  the Hubble function $H$ takes the form:
\begin{equation}\label{hzeq1}
H(z)=\sqrt{\frac{\left[H_0^2 \left(6 \Omega _m+\frac{3
   \varphi _0}{2,5 \cdot 10^6}-6\right)+m \varphi _0\right] \left[V_1+(z+1)^6 \varphi _0^2\right]^{\gamma
   }-(z+1)^3 \left(6 H_0^2 \Omega _m+m \varphi _0\right)}{\left(V_1+\varphi _0^2\right)^{-\gamma } \left[\frac{3 (z+1)^3 \varphi _0}{2,5 \cdot 10^6}-6\right]}}\,.
\end{equation}

It is worth noting that, owing to the presence of an additional parameter, it is also possible include both  $H_0$ and the deceleration parameter $q_0$ in the new parametrization so that the conditions  $H(0)=H_0$ and  $q_0=-H_0 \left(\lim_{z\to 0} \, \frac{d H(z)}{dz}\right)$
allow us to express $V_0$  and $V_1$ in terms of $H_0$, $q_0$ and the other parameters:
\begin{center}
\begin{eqnarray}
&&V_0=2^{-\gamma } \left[6 H_0^2 \left(\Omega _m+2 \varphi _0 \epsilon -1\right)+m \varphi
   _0\right] \left\{\frac{\gamma  \varphi _0^2 \left[6 H_0^2 \left(\Omega _m+2 \varphi _0
   \epsilon -1\right)+m \varphi _0\right]}{H_0^2 \left[6 \Omega _m+q_0 \left(8 \varphi _0
   \epsilon -4\right)+20 \varphi _0 \epsilon -4\right]+m \varphi _0}\right\}^{-\gamma }\,,\nonumber \\ 
   &&V_1=\frac{\varphi _0^2 \left\{2 H_0^2 \left[6 \gamma  \left(\Omega _m+2 \varphi _0 \epsilon
   -1\right)-3 \Omega _m+q_0 \left(2-4 \varphi _0 \epsilon \right)-10 \varphi _0 \epsilon
   +2\right]+(2 \gamma -1) m \varphi _0\right\}}{H_0^2 \left[6 \Omega _m+q_0 \left(8 \varphi
   _0 \epsilon -4\right)+20 \varphi _0 \epsilon -4\right]+m \varphi _0}\,.\nonumber
    \end{eqnarray}
    \end{center}
With these substitutions  the Hubble function $H$ takes the form:

\begin{eqnarray}
&&H(z)=\left\{\frac{1}{12 \varphi _0 (z+1)^3 \epsilon -6}\left[2^{-\gamma } \gamma ^{-\gamma } \left(6 H_0^2 \Omega _m+12 H_0^2 \varphi _0 \epsilon -6  H_0^2+m \varphi _0\right) \right.\right. \times \\
&& \left.\left. \left[6 H_0^2 \left(\Omega _m+2 \varphi _0 \epsilon -1\right)+m
   \varphi _0\right]{}^{-\gamma }  \left\{H_0^2 \left[6 \Omega _m+q_0 \left(8 \varphi _0
   \epsilon -4\right)+20 \varphi _0 \epsilon -4\right]+m \varphi _0\right\}{}^{\gamma } \right.\right.\nonumber \\ &&\left.\left.
   \left\{\displaystyle\frac{2 H_0^2 \left[3 (2 \gamma -1) \Omega _m+2 \left[-3 \gamma +\varphi _0 \epsilon
    \left(6 \gamma -2 q_0-5\right)+q_0+1\right]\right]+(2 \gamma -1) m \varphi _0}{H_0^2
   \left[6 \Omega _m+q_0 \left(8 \varphi _0 \epsilon -4\right)+20 \varphi _0 \epsilon
   -4\right]+m \varphi _0}+(z+1)^6\right\}{}^{\gamma } \right.\right. \nonumber \\ &&\left.\left. \displaystyle-(z+1)^3 \left(6 H_0^2 \Omega _m+m
   \varphi _0\right)\right] \right\}^{\frac{1}{2}}\,.\nonumber
 \end{eqnarray}
\normalsize

In Fig. (\ref{fig1_all}) we compare the  evolution of the Hubble functions  the power law model and in the standard $\Lambda$CDM model.
As we shall see, in order to carry out the comparison with the observational data, it turns out that the above parametrizations are more efficient in exploring the region of parameters. 
\section{Observational data sets}
In our investigation we use  observational data sets on SNIa and GRB Hubble diagram, as well as some recent measurements of $H(z)$ from the differential age of passively evolving elliptical galaxies \cite{farooqb}. We set Gaussian priors on the distance data from the BAO and the Hubble constant $h$. These priors were
included to help break the degeneracies among the parameters of different cosmological models.
\subsection{Supernovae and GRB Hubble diagram}
SNIa observations gave the first strong evidence of the present accelerating
expansion of the Universe \cite{Riess, per+al99}.
Here we consider the recently updated Supernovae Cosmology Project Union 2.1 compilation \cite{Union2.1},
which spans the redshift range $0.015 \le z \le 1.4$.
We compare the theoretically predicted distance modulus $\mu(z)$
with the observed one through a Bayesian approach, based on the definition
of the distance modulus for each of the different models described in the previous section
\begin{equation}
\mu(z_{j}) = 5 \log_{10} ( D_{L}(z_{j}, \{\theta_{i}\}) )+\mu_0\,,
\end{equation}
where $D_{L}(z_{j}, \{\theta_{i}\})$ is the Hubble free luminosity
distance, expressed as a series depending on the cosmological parameters.  The parameter $\mu_{0}$ encodes the Hubble
constant and the absolute magnitude $M$, and has to be
marginalized over. In our analysis we use the version of  the $\chi^2$ given by
\begin{equation}
\label{eq: sn_chi_mod}
\tilde{\chi}^{2}_{\mathrm{SN}}(\{\theta_{i}\}) =  \sum^{{\cal{N}}_{SNIa}}_{j = 1} \frac{(\mu(z_{j}; \mu_{0}=0,
\{\theta_{i})\} -
\mu_{obs}(z_{j}))^{2}}{\sigma^{2}_{\mathrm{\mu},j}}-\frac{\sum^{{\cal{N}}_{SNIa}}_{j = 1} \frac{(\mu(z_{j}; \mu_{0}=0,
\{\theta_{i})\} -
\mu_{obs}(z_{j}))}{\sigma^{2}_{\mathrm{\mu},j}}}{ \sum^{{\cal{N}}_{SNIa}}_{j = 1}
\frac{1}{\sigma^{2}_{\mathrm{\mu},j}}}\,.
\end{equation}

\begin{figure}
\includegraphics[width=\textwidth]{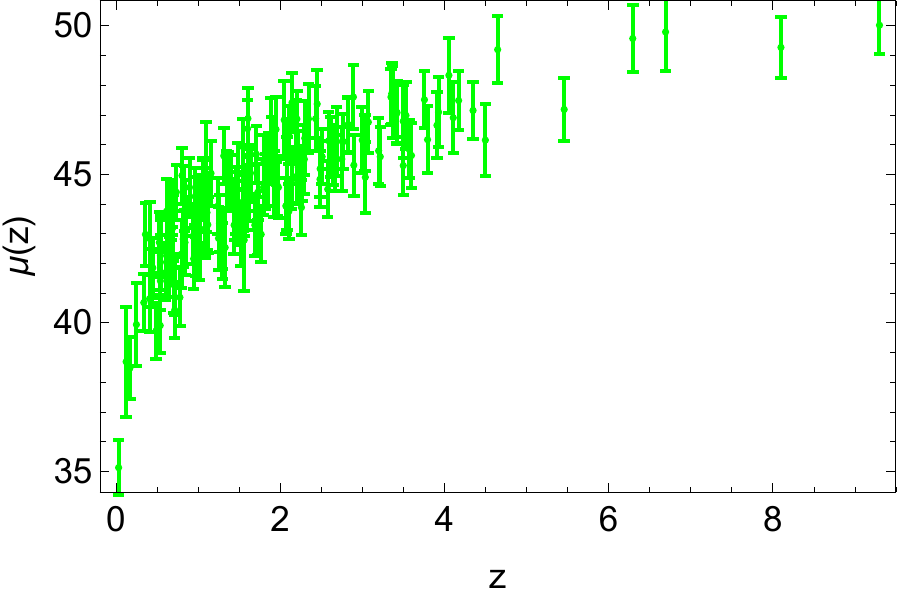}
\caption{GRB Hubble diagram used in our analysis.}
\label{hdreich06}
\end{figure}

Gamma-ray bursts are visible up to very high redshifts thanks to the enormous released amount of energy, and thus are good candidates  for our high-redshift
cosmological investigation. Since their peak luminosity spans a wide range, they are not standard candles; however it is possible to consider them as distance indicators calibrating some empirical correlations of distance-dependent quantities and rest-frame observables  \cite{Amati08}. These
empirical relations allow us to deduce the GRB rest-frame luminosity or energy
from an observer-frame measured quantity, so that the distance modulus can be
obtained with an error that depends essentially on the intrinsic scatter of the
adopted correlation. We perform our analysis using a GRB Hubble diagram data
set, built by calibrating the $E_{\rm p,i}$ -- $E_{\rm iso}$ relation \cite{paper1,paper2}, plotted in Fig. \ref{hdreich06}.

\subsection{Baryon acoustic oscillations and H(z) measurements}
Baryon acoustic oscillations (BAO) data are standard rulers available to investigate
several cosmological models; they are related to density
fluctuations induced by acoustic waves due to primordial
perturbations. Measurements of the Cosmic Microwave Background (CMB) radiation provide the absolute
 scale for these peaks, and the observed position of the peaks
of the two-point correlation function of the  large scale matter distribution enables  to measure distance scales, if it is compared with
these absolute values. In order to use BAOs data, we  define \cite{P10}
\begin{equation}
d_z = \frac{r_s(z_d)}{d_V(z)}\,,
\label{eq: defdz}
\end{equation}
where $z_d$ is the drag redshift computed in \cite{EH98}, $r_s(z)$ is the sound horizon
\begin{equation}
r_s(z) = \frac{c}{\sqrt{3}} \int_{0}^{(1 + z)^{-1}}{\frac{da}{a^2 H(a) \sqrt{1 + (3/4) \Omega_b/\Omega_{\gamma}}}} \ ,
\label{defsoundhor}
\end{equation}
and $d_V(z)$ is the volume distance: 
 \begin{equation}
 d_V(z) = \left[\left(1+z\right) d_A(z)^2\frac{c z}{H(z)}\right]^{\frac{1}{3}}.\label{volumed}
\end{equation}
Here  $ d_A(z)$ is the angular diameter distance. The data used in our analysis are from \cite{Aburg}.
 The measurements of Hubble parameters are a complementary probe to constrain the cosmological parameters and
 investigate dark energy effects \cite{farooqb, farooqa}. The Hubble parameter, defined as the logarithmic derivative of the scale factor
 $H(z) = \displaystyle \frac{\dot a}{a}$,  depends on the differential
 age of the Universe as a function of the redshift and can be measured  using the so-called  cosmic chronometers.
  From spectroscopic surveys we can measure the differential redshift interval $dz$ with high accuracy and  the differential evolution of the age of
  the Universe $dt$ in the  redshift interval $dz$ can be measured too, provided that  optimal probes of the aging of
  the Universe (that is cosmic chronometers) are identified \cite{moresco}. The most reliable cosmic
  chronometers are old early-type galaxies that evolve passively on a timescale much longer than their age difference, which have
formed the most part of  their stars early  and have not experienced further  star formation events. Moreover, the Hubble parameter can also be obtained from the BAO measurements \cite{busca}. We used a list of $28$ $H(z)$ measurements, compiled in \cite{farooqb} and shown in
Table (\ref{hz}).
\begin{table}
\begin{center}
 \centering
  \setlength{\tabcolsep}{1em}
\begin{tabular}{|c|c|c|}
\hline\hline
~~$z$ & ~~$H(z)$ &~~~~~~~ $\sigma_{H}$ \\
~~~~~    & (km s$^{-1}$ Mpc $^{-1}$) &~~~~~~~ (km s$^{-1}$ Mpc $^{-1}$) \\
\hline\hline
0.070&~~        69&~~~~~~~      19.6\\
0.100&~~        69&~~~~~~~      12\\
0.120&~~        68.6&~~~~~~~    26.2\\
0.170&~~        83&~~~~~~~      8\\
0.179&~~        75&~~~~~~~      4\\
0.199&~~        75&~~~~~~~      5\\
0.200&~~        72.9&~~~~~~~    29.6\\
0.270&~~        77&~~~~~~~      14\\
0.280&~~        88.8&~~~~~~~    36.6\\
0.350&~~        76.3&~~~~~~~    5.6\\
0.352&~~        83&~~~~~~~      14\\
0.400&~~        95&~~~~~~~      17\\
0.440&~~        82.6&~~~~~~~    7.8\\
0.480&~~        97&~~~~~~~      62\\
0.593&~~        104&~~~~~~~     13\\
0.600&~~        87.9&~~~~~~~    6.1\\
0.680&~~        92&~~~~~~~      8\\
0.730&~~        97.3&~~~~~~~    7.0\\
0.781&~~        105&~~~~~~~     12\\
0.875&~~        125&~~~~~~~     17\\
0.880&~~        90&~~~~~~~      40\\
0.900&~~        117&~~~~~~~     23\\
1.037&~~        154&~~~~~~~     20\\
1.300&~~        168&~~~~~~~     17\\
1.430&~~        177&~~~~~~~     18\\
1.530&~~        140&~~~~~~~     14\\
1.750&~~        202&~~~~~~~     40\\
2.300&~~        224&~~~~~~~     8\\

\hline\hline
\end{tabular}
\end{center}
\caption{Measurements of the Hubble parameter used in our analysis, as compiled in \cite{farooqb}
}\label{hz}
\end{table}
\section{Statistical analysis}
To constrain the cosmological parameters which characterize the models described in Sect. \ref{models}, we performed a preliminary and standard fitting procedure
to maximize a likelihood function ${\cal{L}}({\bf p})$, in order to identify appropriate starting points for the statistical analysis. This required the knowledge of the precision matrix:

\begin{eqnarray}
\footnotesize
{\cal{L}}({\bf p}) & \propto & \frac {\exp{(-\chi^2_{SNIa/GRB}/2)}}{(2 \pi)^{\frac{{\cal{N}}_{SNIa/GRB}}{2}} |{\bf C}_{SNIa/GRB}|^{1/2}}  \frac{\exp{(-\chi^2_{BAO}/2})}{(2 \pi)^{{\cal{N}}_{BAO}/2} |{\bf C}_{BAO}|^{1/2}} \nonumber\\ ~ & \times & \frac{1}{\sqrt{2 \pi \sigma_{\omega_m}^2}}
 \exp{\left [ - \frac{1}{2} \left ( \frac{\omega_m - \omega^{obs}_{m}}{\sigma_{\omega_m}} \right )^2\right ]} ,\\ &&\times\frac{1}{\sqrt{2 \pi \sigma_h^2}} \exp{\left[ - \frac{1}{2} \left ( \frac{h - h_{obs}}{\sigma_h} \right )^2\right]} \frac{\exp{(-\chi^2_{H}/2})}{(2 \pi)^{{\cal{N}}_{H}/2} |{\bf C}_{H}|^{1/2}}\nonumber
 \\ & \times & \frac{1}{\sqrt{2 \pi \sigma_{{\cal{R}}}^2}} \exp{\left [ - \frac{1}{2} \left ( \frac{{\cal{R}} - {\cal{R}}_{obs}}{\sigma_{{\cal{R}}}} \right )^2 \right ]} \nonumber \,
\label{defchiall}
\end{eqnarray}
where
\begin{equation}
\chi^2(\bf p) = \sum_{i,j=1}^{N} \left( \data_i - x^{th}_i(p)\right)C^{-1}_{ij}  \left( \data_j - x^{th}_j(p)\right) \,.
\label{eq:chisq}
\end{equation}
$\omega_m=h^2\Omega_m$, $\bf p$ indicates the set of the cosmological parameters, $N$ is the number of data,  $\bf \data_i$ is the $i-th$ measurement;
$\bf x^{th}_i(p)$
denote the theoretical predictions for these measurements and depend on the cosmological model and its own parameters $\bf p$;  $\bf C_{ij}$ is
the covariance matrix (specifically,
${\bf C}_{SNIa/GRB/H}$ indicates the SNIa/GRBs/H  covariance matrix);
$(h^{obs}, \sigma_h) = (0.742, 0.036)$\, \cite{shoes},
and $(\omega_{m}^{obs}, \sigma_{\omega_m}) = (0.1356, 0.0034)$\, \cite{PlanckXIII}.
We tested that  our results are not biased by the choice of the  prior on h. 
The term
$\displaystyle  \frac{1}{\sqrt{2 \pi \sigma_{{\cal{R}}}^2}} \exp{\left [ - \frac{1}{2} \left ( \frac{{\cal{R}} - {\cal{R}}_{obs}}{\sigma_{{\cal{R}}}} \right )^2 \right ]} $
in the likelihood (\ref{defchiall}) takes into account the so called shift parameter ${\cal{ R}}$, defined as

\begin{equation}
{\cal{R}} = H_{0} \sqrt{\Omega_M} \int_{0}^{z_{\star}}{\frac{dz'}{H(z')}}\,,
\label{eq: defshiftpar}
\end{equation}
being  $z_\star = 1090.10$   \cite{B97}, \cite{EB99}. According to
the Planck data  $({\cal{R}}_{obs}, \sigma_{{\cal{R}}}) = ( 1.7407, 0.0094)$.

\begin{figure}
\includegraphics[width=\textwidth]{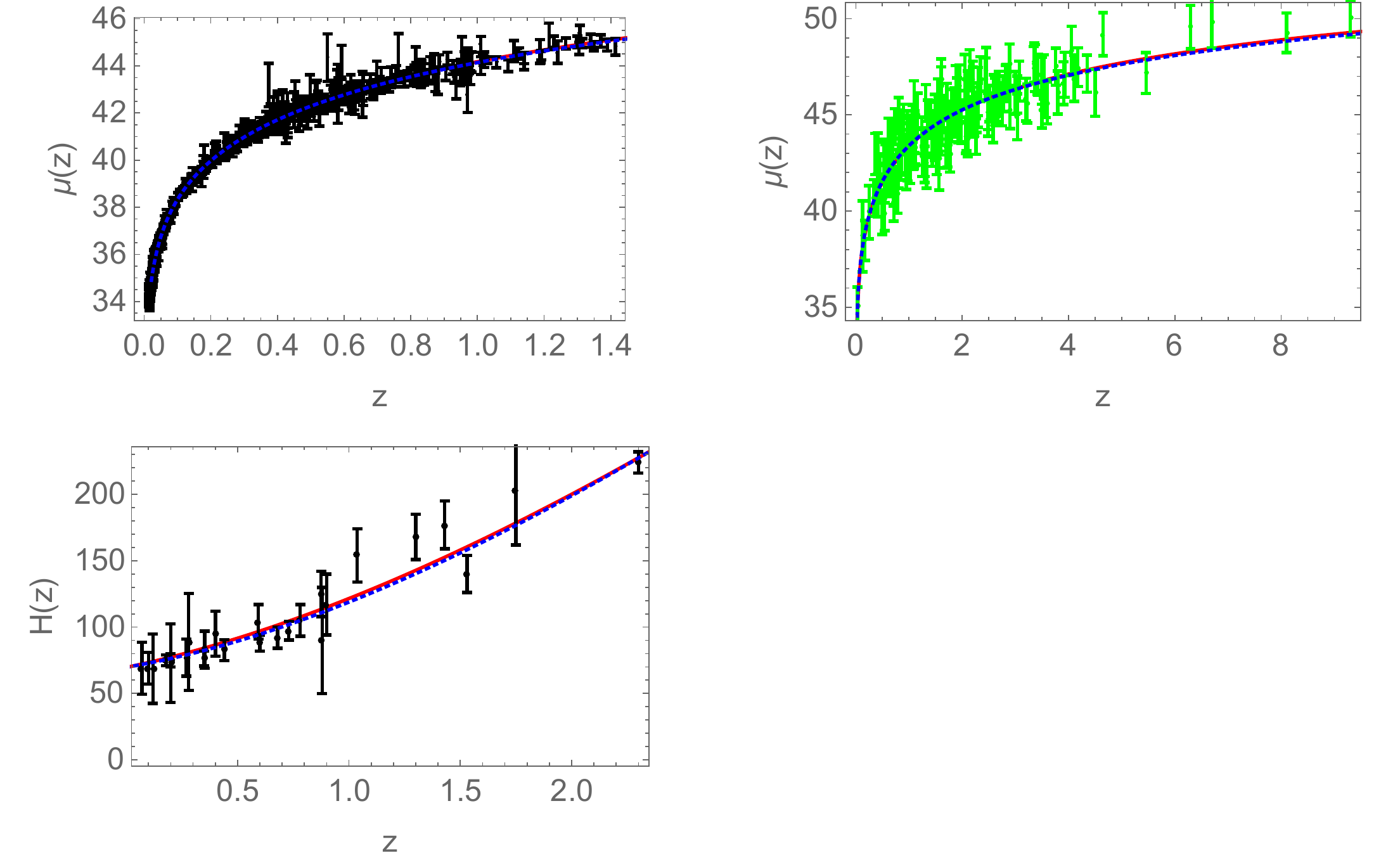}
\caption{ Comparison between the observational data and the theoretical predictions  for the power law potential  (red solid line), and the  exponential potential, corresponding to their own best fit values for the parameters.}
\label{fitdata}
\end{figure}

To  sample the ${\cal{N}} $ dimensional space of
parameters corresponding to each of our models,  we used the  MCMC method and ran five parallel chains, the convergence of which has been tested using 
the reduction factor ${R^*}$, defined as the square root of the ratio of
the variance extra-chain and the variance intra-chain.  A large ${R^*}$
indicates that the extra-chain variance is substantially greater than the
intra-chain variance, so that a longer simulation is needed. We required that the convergence was reached if 
${R^*}$  approached 1 for each parameter:  we set the precision of order $0.05$.  As first
step, which allowed us to select the starting points of the full analysis, we ran our chains to
compute the likelihood in Eq. (\ref{defchiall}) considering only the SNIa data.
Therefore we applied  the same  MCMC approach to evaluate the likelihood in Eq.
(\ref{defchiall}), combining all the data,
as described above. We discarded the  first $30\%$  of the point iterations at
the beginning of any MCMC run, and thinned the chains that were run many times.
We finally   extracted the constraints on the parameters by coadding the
thinned chains.  The histograms of the parameters from the merged chains were
then used to infer median values and confidence ranges: the $15.87$th and
$84.13$th quantiles define the $68 \%$ confidence interval; the $2.28$th and
$97.72$th quantiles define the $95\%$ confidence interval; the $0.13$th and
$99.87$th quantiles define the $99\%$ confidence interval. In Table \ref{tabcosmofit},
we presented the results of our analysis: we indicated the mean, the median, the $1\sigma$ and $2\sigma$ regions of confidence for the main parameters of  the different models.
In Fig. (\ref{fitdata}) we plot the observational data with the best fit model relative to the power law and exponential potential.
\begin{table}
\begin{center}
\resizebox{8 cm}{!}{
\begin{tabular}{cccccc}
\hline
$Id$ & $\langle x \rangle$ & $\tilde{x}$ & $68\% \ {\rm CL}$  & $95\% \ {\rm CL}$ \\
\hline \hline\\
~ & \multicolumn{3}{c}{$V(\varphi)=V_0 \varphi^\alpha$} & \multicolumn{2}{c}{ }\\
\hline
~ & ~ & ~ & ~ & ~ & ~  \\
$\Omega_m$ & 0.21 &0.22& (0.19,\, 0.24) & (0.17,\, 0.25) \\
~ & ~ & ~ & ~ & ~ & ~  \\
$h$ &  0.69& 0.70 & (0.67,\, 0.71) & (0.66,\, 0.72) \\
~ & ~ & ~ & ~ & ~ & ~  \\
$\alpha$ &0.23&0.21 & (0.19,\,0.24) & (0.15,\, 0.30)  \\
~ & ~ & ~ & ~ & ~ & ~  \\
$m$ &1.56 &1.60& (1.39,\,1.7) & (1.31,\,1.76)  \\
~ & ~ & ~ & ~ & ~ & ~   \\
\hline\hline\\
~ & \multicolumn{3}{c}{$V(\varphi)=V_0 \exp(-\lambda \varphi) $} & \multicolumn{2}{c}{}\\
\hline
~ & ~ & ~ & ~ & ~ & ~  \\
$\Omega_m$ & 0.25 &0.25& (0.23,\, 0.27) & (0.21,\, 0.29) \\
~ & ~ & ~ & ~ & ~ & ~  \\
$h$ &  0.69 & 0.69 & (0.69,\, 0.71) & (0.68,\, 0.72) \\
~ & ~ & ~ & ~ & ~ & ~  \\
$\lambda$ &0.67&0.68 & (0.25,\,1.1) & (0.21,\, 1.35)  \\
~ & ~ & ~ & ~ & ~ & ~  \\
$m$ &0.33 &0.24& (0.1,\,0.78) & (0.06,\,0.96)  \\
~ & ~ & ~ & ~ & ~ & ~   \\
\hline\hline\\
~ & \multicolumn{3}{c}{$V(\varphi)=V_0 (\varphi^{2}+V_1)^{\gamma}$} & \multicolumn{2}{c}{}\\
\hline
~ & ~ & ~ & ~ & ~ & ~  \\
$\Omega_m$ & 0.21 &0.23 & (0.20,\, 0.24) & (0.16,\, 0.27)  \\
~ & ~ & ~ & ~ & ~ & ~  \\
$h$ &  0.70& 0.71 & (0.67,\, 0.73) & (0.65,\, 0.74) \\
~ & ~ & ~ & ~ & ~ & ~  \\
$q_{0}$ &-0.66& -0.63 & (-0.65,\,-0.57) & (-0.67,\, -0.49)\\
~ & ~ & ~ & ~ & ~ & ~  \\
$\gamma$ &0.51& 0.6 & (0.47,\,0.91) & (0.45,\, 1.05) \\
~ & ~ & ~ & ~ & ~ & ~  \\
$m$ &0.87& 0.8 & (0.66,\,0.95) & (0.52,\, 1.17) \\
~ & ~ & ~ & ~ & ~ & ~  \\
\hline\hline\\
\end{tabular}}
\end{center}
\caption{Constraints on the main cosmological parameters which enter in the representation of the models described in Sect. \ref{models}. The likelihood has been marginalized with respect the others.  Columns report the mean $\langle x \rangle$ and median $\tilde{x}$ values  and the $68\%$ and $95\%$ confidence limits.  }
\label{tabcosmofit}
\end{table}

\section{ Comparison of our fermionic models with the Chevallier-Polarski-Linder (CPL) model }
\begin{figure}
\includegraphics[width=\textwidth]{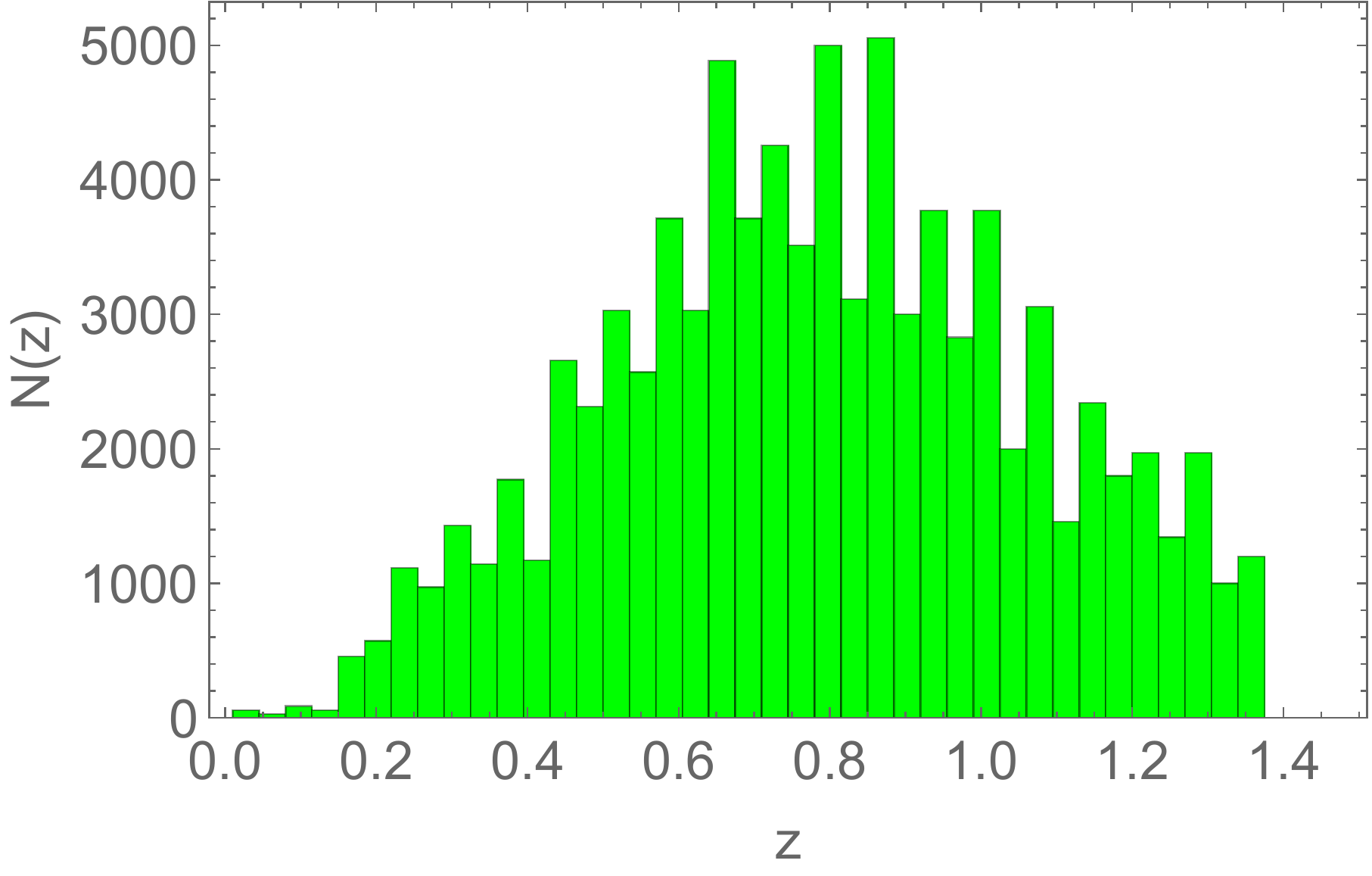}
\caption{Simulated data for an Euclid-like mission: we plot the SNeIa redshift distribution used in our analysis.}
\label{Euclidlikenz}
\end{figure}

\begin{figure}
\includegraphics[width=\textwidth]{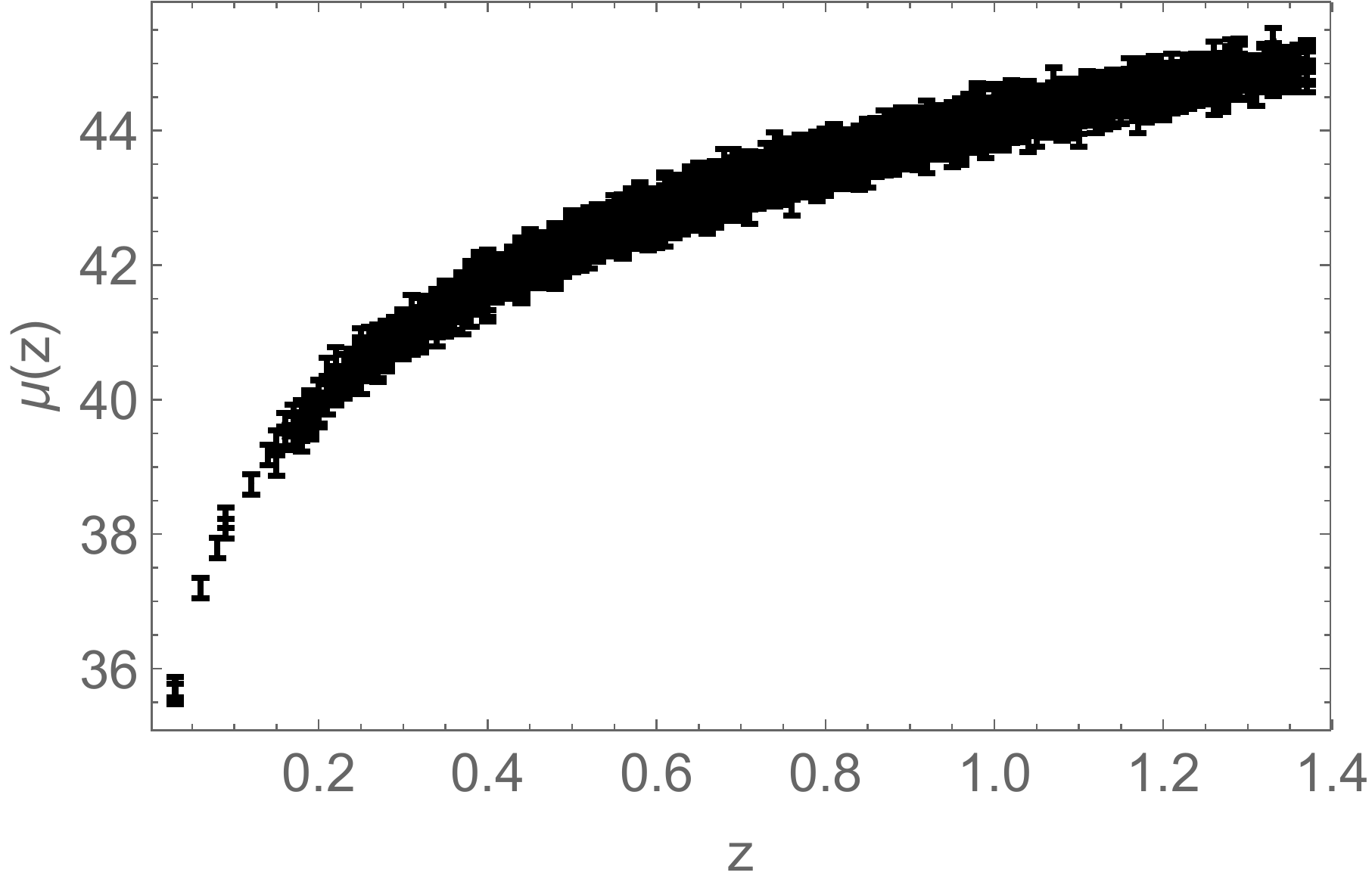}
\caption{ The mock dataset used in our analysis.}
\label{SimSN}
\end{figure}

  In this section we compare the different models presented in the previous sections and check if we can discriminate 
against them. We use the Akaike Information Criterion (AIC) \cite{aic,aic2}, and its indicator
\begin{equation} \label{aic}
AIC=-2 \ln{\cal{L}_{\bf max} }+  2 k_p +\frac{2 k_p (k_p+1)}{N_{tot}-k_p-1}\,,
\end{equation}
where $N_{tot}$ is the  total number of data and $k_p$ the number of free parameters (of the cosmological model) . 
 It turns out that the smaller is the value of AIC the better is  the fit to  the data. To compare different cosmological models  we introduce the model  difference
$ \Delta_{AIC} = AIC_{model} - AIC_{min}$. The relative difference corresponds to different cases:  
$4 < \Delta_{AIC} < 7$ indicates a positive evidence against the model with higher value of $AIC_{model}$, 
while $\Delta_{AIC} \geq 10 $ indicates a strong evidence. $ \Delta_{AIC} \leq 2$ is an indication that 
the two  models are consistent. In our case we have found that the model with the lower AIC is the exponential potential model $V(\varphi)=V_0 \exp\left(-\lambda\varphi\right)$. It turns out that $ \Delta_{AIC} = 4.3 $ if we consider the exponential  potential and  $ \Delta_{AIC} = 1.97 $ for the the extended power law potential. We can use he same method to make a comparison of our model with the CPL parametrization for dark energy, which assumes a dark energy EOS given by
\begin{equation}
w(z) =w_0 + w_{1} z (1 + z)^{-1} \,,
\label{cpleos}
\end{equation}
where $w_0$ and $w_{1}$ are real numbers that represent the EOS present value and its overall time evolution,
respectively \cite{cpl1,cpl2}. For high redshift we have the following behavior
\begin{equation}
\lim_{z \to \infty}w(z)=w_0+w_{1}\,.
\end{equation}
If we compare the exponential potential model with the CPL, it turns out that the exponential model has got the lower AIC, and 
 $\Delta_{AIC} = 5.9 $. This indicates a weak evidence against the CPL model. To confirm the results obtained above we rely on future data. Here we investigate the possibility to constrain our model using simulated data from an Euclid\,-\,like survey \cite{Euclid}. The number of SNeIa which could be used for cosmology and their redshift distribution is plotted in Fig.\,\ref{Euclidlikenz}. 
To each simulated SN  we estimate the error on the distance modulus as \cite{kim}\,:

\begin{equation}
\sigma_{\mu}(z) = \sqrt{\sigma_{sys}^2 + (z/z_{max})^2 \sigma_m^2}\,.
\label{eq: sigmamusn}
\end{equation}
Here $z_{max}$ is the maximum redshift of the sample, $\sigma_{sys}$ an instrinsic scatter and $\sigma_m$ depends on the photometric accuracy. In our case  $(z_{max}, \sigma_{sys}, \sigma_m) = (1.4, 0.15, 0.02)$. We then assign to each SN a distance modulus randomly generated from a Gaussian distribution centered on a fiducial model $\mu_{fid}(z)$ and variance $\sigma_{\mu}(z)$.  In this analysis we set the exponential potential model as the fiducial model. In Fig. \ref{SimSN} we plot the mock dataset. 
In Table \ref{SimSNexp} we summarized the results of the simulation, when we consider the simulated SneIa HD  as cosmological probes: it turns out that the mock dataset is able to constraint much better our model.
\begin{table}
\begin{center}
\resizebox{8 cm}{!}{
\begin{tabular}{cccccc}
\hline
$Id$ & $\langle x \rangle$ & $\tilde{x}$ & $68\% \ {\rm CL}$  & $95\% \ {\rm CL}$ \\
\hline \hline\\
\hline
~ & ~ & ~ & ~ & ~ & ~  \\
$\Omega_m$ & 0.27 &0.27 & (0.26,\, 0.27) & (0.25,\, 0.28)  \\
~ & ~ & ~ & ~ & ~ & ~  \\
$H_0$ &  70.1& 70.1 & (69.8,\, 70.5) & (69.5,\, 70.7) \\
~ & ~ & ~ & ~ & ~ & ~  \\
$\varphi_{0}$ &0.005& 0.005 & (0.004,\,0.006) & (0.003,\, 0.007)\\
~ & ~ & ~ & ~ & ~ & ~  \\
$m$ &0.60& 0.51 & (0.39,\,0.89) & (0.30,\, 0.99) \\
~ & ~ & ~ & ~ & ~ & ~  \\
$\lambda$ &2.45& 2.50 & (2.04,\,2.87) & (1.54,\, 2.97) \\
~ & ~ & ~ & ~ & ~ & ~  \\
\hline\hline\\
\end{tabular}}
\end{center}
\caption{Results of our statistical analysis, when we consider the simulated SneIa HD  as cosmological probes. Columns report the mean $\langle x \rangle$ and median $\tilde{x}$ values  and the $68\%$ and $95\%$ confidence limits. }
\label{SimSNexp}
\end{table}

\section{A new dynamical system formulation}\label{dynamical}
We can combine the data analysis above with the dynamical system approach to infer further features of the cosmic history that corresponds to the  observational results we have obtained.

Under the condition $ \varepsilon>0$ and $\varphi_0>0$, we can consider the dynamical system variables
\begin{equation}\label{DynSysVarC}
\begin{split}
\Omega =\sqrt{\frac{\rho }{\mathcal{D}^2}},\qquad Y=\sqrt{\frac{m
   \varphi }{2\mathcal{D}^2}}, \\
E=\sqrt{\frac{2 \epsilon \dot{\varphi}^2}{3 \varphi\mathcal{D}^2}}, \qquad Q=\frac{H}{\mathcal{D}},
 \end{split}
\end{equation} 
where
\begin{equation}
\mathcal{D}^2=3 H^2+\frac{V(\varphi )}{2}.
\end{equation}
This choice sends fixed points for which $\mathcal{D}=0$ to the asymptotic part of the phase space. Because of eq.~\eqref{14.1} this condition reduces to the differential equation 
\begin{equation}
\frac{\dot{\varphi}^2}{3\varphi^2}+\frac{V(\varphi )}{2}=0
\end{equation}
which, given the form of the potential, represents the solution associated to the asymptotic limit of the phase space. 
Using variables \eqref{DynSysVarC}, the cosmological equations are equivalent to the dynamical system
\begin{align}
\Omega_{,\mathcal{N}}&=-\frac{3 Q \Omega}{2 \left(E^2-3 Q^2\right)}  \left[E^2 \left(3 Q^2-1\right)
  \mathds{V}+E^2 (w+1)+3 w Q^2  \left(\Omega^2-1\right)\right],\\
   E_{,\mathcal{N}}&=-\frac{3 E Q}{2 \left(E^2-3 Q^2\right)}
   \left[\left(E^2-1\right) \left(3 Q^2-1\right)
  \mathds{V}+E^2+w \Omega ^2 \left(3 Q^2-1\right)-1\right],\\
 Q_{,\mathcal{N}}&=-\frac{3 Q^2 }{2
   \left(E^2-3 Q^2\right)}\left(3
   Q^2-1\right) \left[\left(E^2-1\right)
 \mathds{V}+w \Omega ^2+1\right],
\end{align}
together with the decoupled equation 
\begin{equation}
\mathcal{D}_{,\mathcal{N}}=\frac{3 \mathcal{D}^2 Q}{2 \left(E^2-3 Q^2\right)} \left[E^2
\left(3 Q^2-1\right) \mathds{V}+3 Q^2 \left(w\Omega ^2+1\right)\right],
\end{equation}
and the constraint
\begin{equation}\label{FriedConstrDS}
E^2+Y^2+\Omega^2=1
\end{equation}
where we have defined 
\begin{equation}
X_{,\mathcal{N}}= \frac{1}{\mathcal{D}}\dot{X}
\end{equation}
and 
\begin{equation}
\mathds{V}=\mathds{V}\left(\frac{E^2}{ 6 \varepsilon Q^2}\right)=\left.\frac{\varphi  V'(\varphi )}{V(\varphi )}\right|_{\varphi=E^2/ 6 \varepsilon Q^2}
\end{equation}
The constraint \eqref{FriedConstrDS} implies that the phase space is compact. 

In the following, however, we will consider  the case $w=0$.  For this value of the barotropic factor the  cosmological equations \eqref{FinalEqCosm2} imply that $\rho=\beta \varphi$, where $\beta=\rho_0/\varphi_0$ is a constant. This means that the phase space will loose one dimensions. Therefore, if $w=0$, we can choose the variables
\begin{equation}
\begin{split}
 \bar{Y}=\sqrt{\frac{(m+2\beta)\varphi }{2\mathcal{D}^2}}, \qquad E=\sqrt{\frac{2 \epsilon \dot{\varphi}^2}{3 \varphi\mathcal{D}^2}}, \qquad Q=\frac{H}{\mathcal{D}},
 \end{split}
\end{equation}  
so that the dynamical equations can be written as 
\begin{align}
  E_{,\mathcal{N}}&=-\frac{3 E Q \left(E^2-1\right) \left[\left(3 Q^2-1\right)
   \mathbb{V}+1\right]}{2 \left(E^2-3 Q^2\right)},\\
 Q_{,\mathcal{N}}&=-\frac{3 Q^2 \left(3
   Q^2-1\right) \left[\left(E^2-1\right) \mathbb{V}+1\right]}{2 \left(E^2-3
   Q^2\right)}.
\end{align}
with the constraint
\begin{equation}\label{vincolo2}
 \bar{Y}^2=1-E^2.
\end{equation}
The above system is still not compact in the variable $Q$. We can obtain a compact two-dimensional system with the transformation
\begin{equation}
E=\cos\theta,\qquad \bar{Y}=\sin\theta,\qquad Q=\frac{\bar{Q}}{1-\bar{Q}},
\end{equation} 
which leads to the system
\begin{align}
  \bar{Q}_{,\mathcal{N}}&=\frac{3 \bar{Q}^2 \left[2 \bar{Q} \left(\bar{Q}+1\right)-1\right]
   \left(\bar{\mathbb{V}} \sin ^2\theta -1\right)}{2 \left(\bar{Q}-1\right)^2 \cos
   ^2\theta -6 \bar{Q}^2},\\
 \theta_{,\mathcal{N}}&=\frac{3 \bar{Q} \sin \theta \cos \theta 
   \left\{\left[2 \bar{Q} \left(\bar{Q}+1\right)-1\right]
   \bar{\mathbb{V}}+\left(\bar{Q}-1\right)^2\right\}}{2 \left(\bar{Q}-1\right)[ \left(\bar{Q}-1\right)^2 \cos
   ^2\theta -3  \bar{Q}^2]},
\end{align}
with the decoupled equation
\begin{equation}
\mathcal{D}_{,\mathcal{N}}=\frac{3 Q
   \mathcal{D}^2 \left[E^2 \left(3 Q^2-1\right) \bar{\mathbb{V}}+3
   Q^2\right]}{2 \left(E^2-3 Q^2\right)},
\end{equation}
where
\begin{equation}
\bar{\mathds{V}}=\mathds{V}\left(\frac{\left(\bar{Q}-1\right)^2 \cos ^2\theta }{\bar{Q}^2}\right).
\end{equation}
We will now use the data analysis above to determine the point of the phase space which corresponds to the current state of the universe. Using this result and the dynamical system above we will deduce, for each type of potential, information on the entire cosmic history that is associated with the observations.

\subsection{The case $V(\varphi)=V_0 \varphi^\alpha$ }\label{PowerV1}
If one considers the power law potential $V(\varphi)=V_0 \varphi^\alpha$, then
\begin{equation}\label{DynSysPowV1}
\begin{split}
  \bar{Q}_{,\mathcal{N}}&=\frac{3 \bar{Q}^2 \left[2 \bar{Q} \left(\bar{Q}+1\right)-1\right]
   \left(\alpha \sin ^2\theta -1\right)}{2 \left(\bar{Q}-1\right)^2 \cos
   ^2\theta -6 \bar{Q}^2},\\
 \theta_{,\mathcal{N}}&=\frac{3 \bar{Q} \sin \theta \cos \theta 
   \left\{\left[2 \bar{Q} \left(\bar{Q}+1\right)-1\right]
   \alpha+\left(\bar{Q}-1\right)^2\right\}}{2 \left(\bar{Q}-1\right)[ \left(\bar{Q}-1\right)^2 \cos
   ^2\theta -3  \bar{Q}^2]},
\end{split}
\end{equation}
and 
\begin{equation}
\mathcal{D}_{,\mathcal{N}}=\frac{3 \bar{Q}
   \mathcal{D}^2 \left[E^2 \left(3 \bar{Q}^2-1\right) \alpha+3
   \bar{Q}^2\right]}{2 \left(E^2-3 \bar{Q}^2\right)},
\end{equation}
The system \eqref{DynSysPowV1} has invariant submanifolds at $\bar{Q}=0$, $\bar{Q}=\frac{1}{2} \left(\sqrt{3}-1\right)$, $\theta=k\pi$, $\theta=\pi/2 + k\pi$ with $k$ integer number. It is also singular in 
\begin{equation}
\bar{Q}=\frac{\cos (2 \theta )\pm2 \sqrt{3} \cos \theta +1}{\cos (2 \theta )-5}\qquad \bar{Q}=1
\end{equation}
However, as we have seen, the divergence in $\bar{Q}=1$ is just an artifact of our variable choice.  

Setting $(\bar{Q}_{,\mathcal{N}}=0, \theta_{,\mathcal{N}}=0)$ shows the presence of two physical fixed points in $[0,2\pi]$ and a fixed line. The stability of these points and their associated solutions are given in Table \ref{TableV1}.  The line $\mathcal{L}$ is composed of non hyperbolic fixed points and one should calculate the Centre manifold for these points. However, since the total dimension of the phase space is two, the stability of these points can be calculated simply taking the second derivative of the equation for $\bar{Q}$ at second order at $\bar{Q}=0$. It turns out that for 
\begin{align}
&\alpha >1, \quad k\pi<\theta< (2k+1)\frac{\pi}{2} \\ 
&\alpha <1, \quad (2k+1)\frac{\pi}{2}<\theta< k\pi
\end{align}  
where $k\in \mathds{Z}$, these points are unstable. Since on the line $\mathcal{L}$ fixed points can lead to unstable static solutions, we can conclude that these cosmologies admit bounces and turning points only for certain values of $\alpha$.

Comparing with the work in \cite{CVC}, we see that there are some differences; in particular, some of the fixed points do not appear in the compact version of the phase space. This is due to two different occurrences. The first is that in these points the function $\mathcal{D}$ is either zero or divergent and in this case there is no connection between the different variable systems. In the case of the power law potential we have
\begin{equation}
\mathcal{D}=0\Rightarrow \dot{\varphi}^2+V_0\frac{3}{2}\varphi^{\alpha+2}=0
\end{equation}
which implies via eq. \eqref{14.1}
\begin{equation}\label{AttV1}
a=a_0(t-t_0)^{\frac{2}{3\alpha}}.
\end{equation}
if $V_0$ is negative. This implies that for $V_0<0$ orbits approaching the asymptotic border will represent a Universe whose expansion approaches to the evolution above. 
 
The second occurrence is that some of these points correspond to special cases in which two of the coordinates are one divergent and the other complex divergent. In this case the constraint \eqref{FriedConstrDS} is still satisfied. However, in the new formulation we have assumed that $\varphi$ is positive, whereas these points are compatible with a negative $\varphi$. We can say therefore that they have no relevance if $\varphi_0>0$ and that our compact formulation catches the true degrees of freedom of the cosmological model. Similar reasonings can be made for the other models we will consider here.

\begin{figure}
\centering
\includegraphics[width=0.8\textwidth]{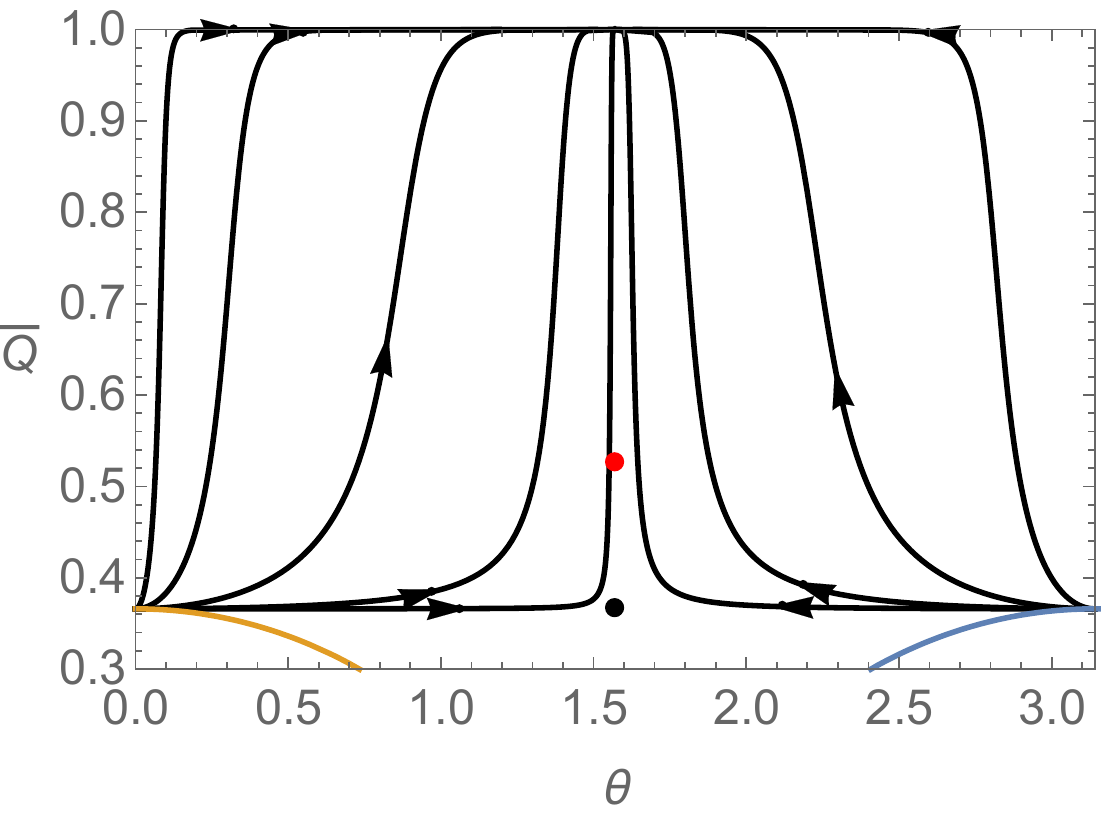}
\caption{Portion of the phase space of equations \eqref{DynSysPowV1}. Here the black dot represents the fixed point $\mathcal{A}_1$ and the blue and blue lines the singularity of the dynamical system. The red dot represents the state of the universe as indicated from the observational analysis of Sections 4, 5 and 6.}
\label{figPSV1}
\end{figure}

\begin{figure}
\centering
\includegraphics[width=0.8\textwidth]{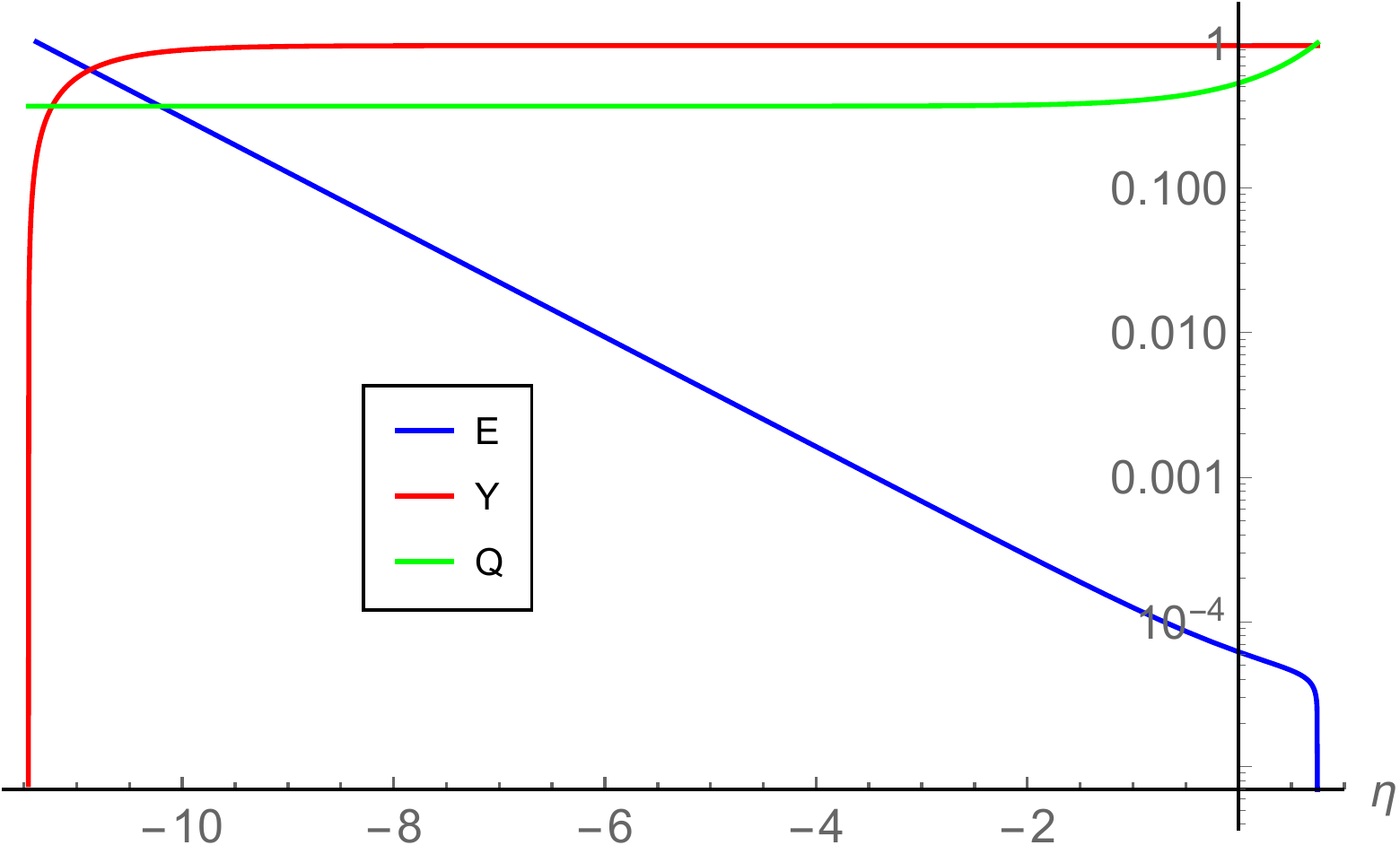}
\caption{Semilogarithmic plot of the evolution of the dynamical system variables along the orbit associated with the conditions \eqref{ICV1}. }
\label{figDynVarV1}
\end{figure}

\begin{figure}
\centering
\includegraphics[width=0.8\textwidth]{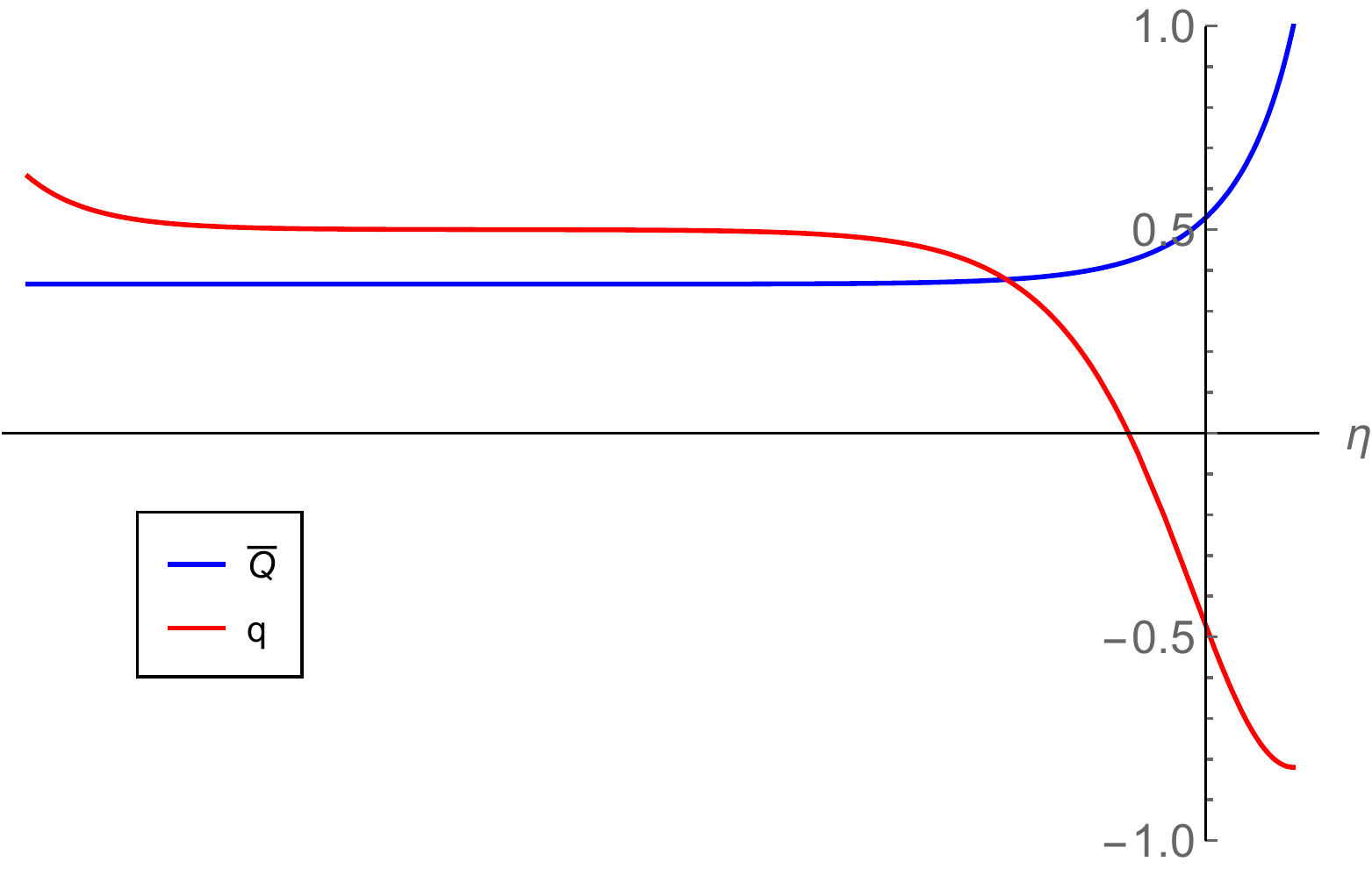}
\caption{Plot of the evolution of the variable $\bar{Q}$ and the deceleration factor $q$ along the orbits of the phase space for the potential  $V(\varphi)=V_0 \varphi^\alpha$ associated with the conditions \eqref{ICV1}. }
\label{figQqV1}
\end{figure}

The data analysis above indicates the following values of the quantities appearing in the cosmological equations at present\footnote{In order to correctly evaluate the magnitude of the fitted values in the SI units, it is worth noting, here and in the cases of the other potentials,  that we are using natural units.  Moreover,  the Hubble constant  is usually estimated as  $H_0=100 h  \frac{Km}{Mpc} s^{-1}$ , and the conversion factor from $Km$ to $Mpc$ is  $\zeta\simeq 3.24\, 10^{-20}$. Actually, for $h=0.696$ the actual dark matter density is  $\rho_0=2.41\, 10^{-27} \,Kg/m^3$, as expected. }:  
\begin{equation}\label{ICV1}
\begin{split}
&\varepsilon \rightarrow 10^{-7},\quad H_0 \rightarrow  69.6 ,\quad  \rho_0 \rightarrow 3851.11 ,\quad \varphi _0 \rightarrow 0.006 ,\\ &V_0 \rightarrow -40344.1 ,\quad \alpha \rightarrow  0.12 ,\quad m \rightarrow 1.8
\end{split}
\end{equation}
In Fig. \ref{figPSV1} we have represented the compact phase space and the position representing the state of the Universe related to the data analysis of the previous sections. It is evident that the past attractor of the dynamics is a singular state reached coasting the invariant submanifold  $\bar{Q}=\frac{1}{2} \left(\sqrt{3}-1\right)$ and the Universe approaches the $\bar{Q}=1$ boundary after transiting close the fixed point $\mathcal{A}_1$. This behavior can be observed directly plotting the behavior of the dynamical variables in time (see Fig. \ref{figDynVarV1}). Indeed, plotting the behavior of the deceleration factor $q$ (see Fig. \ref{figQqV1}) one can conclude that, as expected, the observational data represent a cosmology which present cosmic acceleration and determine the value of $q$ on the boundary. This result is consistent with the interpretation we have given of the $\bar{Q}=1$ boundary. In fact the solution \eqref{AttV1} for the value of $\alpha$ given in \eqref{ICV1} corresponds to an accelerated solution. The solution \eqref{AttV1} was also an attractor in the phase space analysis of \cite{CVC}, but it was not obvious that the initial conditions \eqref{ICV1} would lead to it. 

\begin{table}[tbp] \centering
\caption{The fixed points and the solutions of the NMC model with dust matter and $V=V_0 \varphi^{\alpha}$. Here $S$ stays for saddle, $A$ for attractor and $R$  for repeller.}
\begin{tabular}{cclllccc}
& & \\
\hline   Point &$[\theta,\bar{Q}]$ &  Scale Factor &  Stability  \\ \hline
& & \\
$\mathcal{A}_k$ &$\left[\frac{\pi}{2}+k\pi,\frac{1}{2}\left(1-\frac{1}{\sqrt{3}}\right)\right]$& $a=a_0\left(t-t_0\right)^{2/3}$&  S\\
& & \\
\multirow{4}{*}{$\mathcal{L}$}& \multirow{4}{*}{$\left(\theta_*,0\right)$}& \multirow{4}{*}{$a=a_0 t^{\frac{2 \cos\theta_*}{3 \left(\alpha -1\right)}}$} & R $\alpha >1, \quad k\pi<\theta< (2k+1)\frac{\pi}{2}$ \\
& & & R $\alpha <1, \quad (2k+1)\frac{\pi}{2}<\theta< k\pi$\\
& & & A $\alpha >1, \quad (2k+1)\frac{\pi}{2}<\theta< k\pi$\\
& & & A $\alpha <1, \quad k\pi<\theta< (2k+1)\frac{\pi}{2}$\\ 
& & \\
 \hline
\end{tabular}\label{TableV1}
\end{table}

\subsection{The case $V(\varphi)=V_0 \exp\left(-\lambda\varphi\right)$ }\label{ExpV2}
\begin{figure}
\centering
\includegraphics[width=0.8\textwidth]{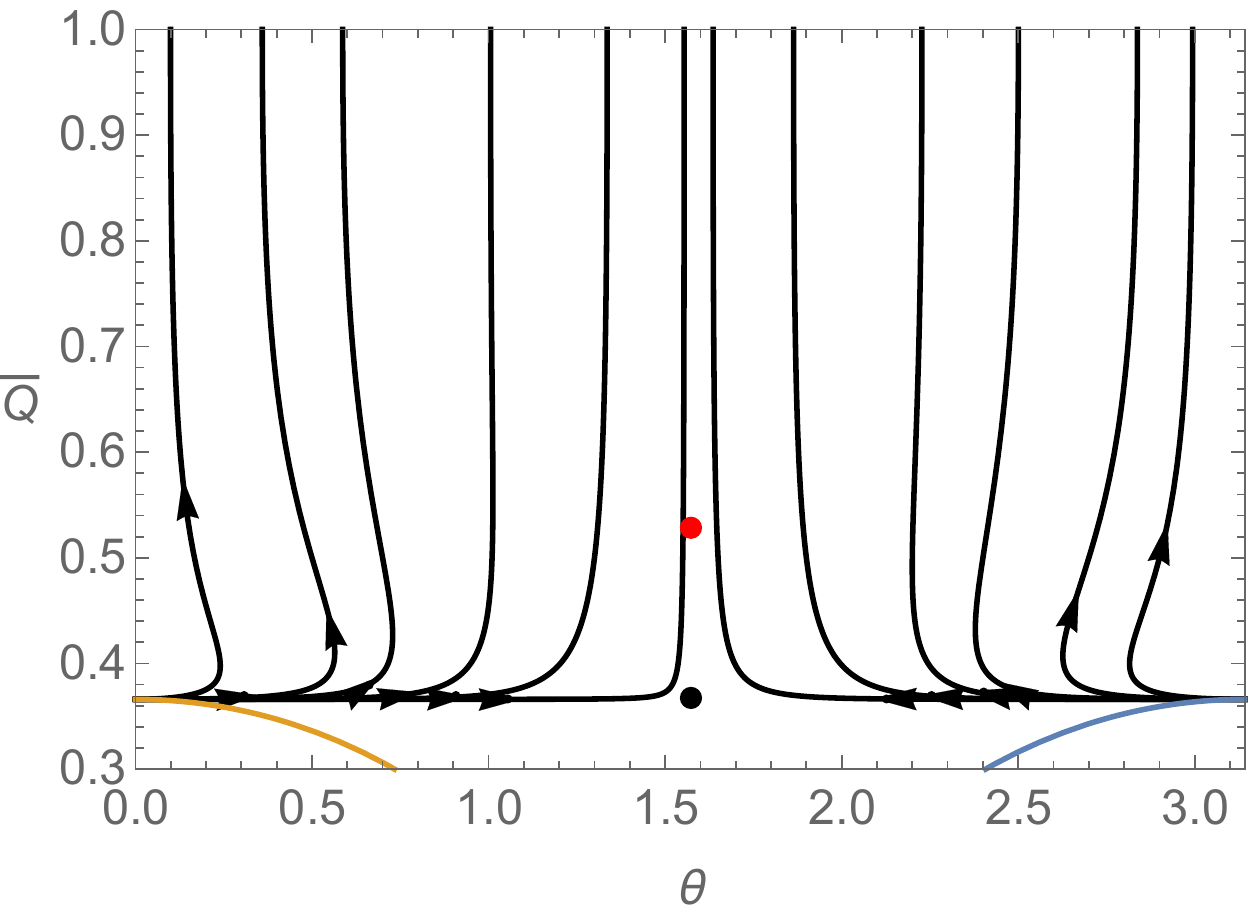}
\caption{Portion of the phase space of equations \eqref{DynSysExpV2}. Here the black dot represents the fixed point $\mathcal{A}_1$ and the blue and blue lines the singularity of the dynamical system. The red dot represents the state of the universe as indicated from the observational analysis of Sections 4, 5 and 6.}
\label{figPSV2}
\end{figure}

\begin{figure}
\centering
\includegraphics[width=0.8\textwidth]{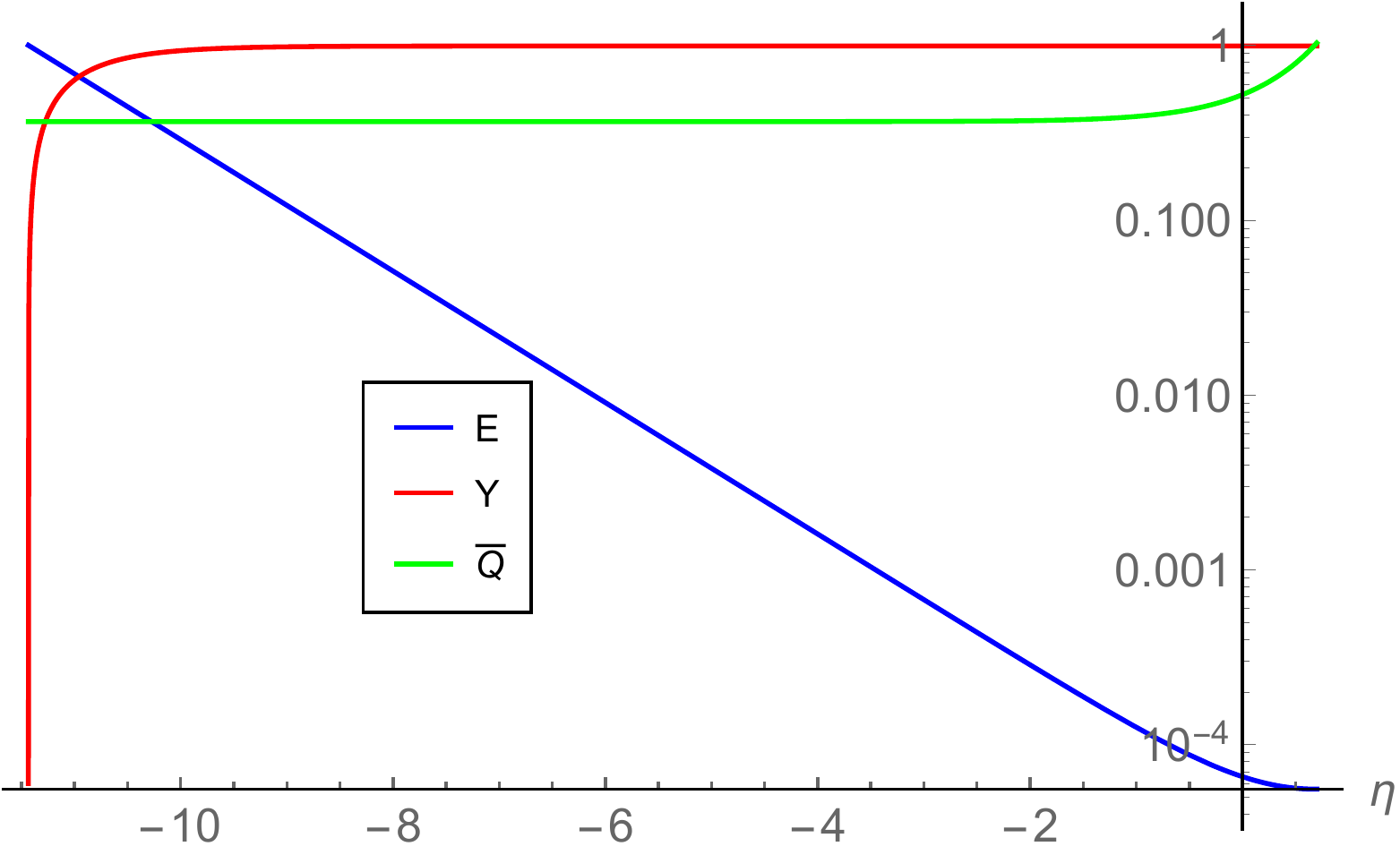}
\caption{Semilogarithmic plot of the evolution of the dynamical system variables along the orbit associated with the conditions \eqref{ICV2}. }
\label{figDynVarV2}
\end{figure}

If one consider the exponential  potential $V(\varphi)=V_0 \exp\left(-\lambda\varphi\right)$, then
\begin{equation}\label{DynSysExpV2}
\begin{split}
  \bar{Q}_{,\mathcal{N}}&=\frac{3  \left[1-2 \bar{Q} \left(\bar{Q}+1\right)\right]
   \left[\lambda  (\bar{Q}-1)^2 \cos ^2(\theta )\sin ^2\theta  +\bar{Q}^2\right]}{2 \left(\bar{Q}-1\right)^2 \cos
   ^2\theta -6 \bar{Q}^2},\\
 \theta_{,\mathcal{N}}&=\frac{3  \sin \theta \cos \theta 
   \left\{\lambda  (\bar{Q}-1)^2 \left[1-2 \bar{Q} \left(\bar{Q}+1\right)\right]\cos ^2(\theta) 
+\bar{Q}^2\left(\bar{Q}-1\right)^2\right\}}{2 \bar{Q}\left(\bar{Q}-1\right)[ \left(\bar{Q}-1\right)^2 \cos^2\theta -3  \bar{Q}^2]},
\end{split}
\end{equation}
and 
\begin{equation}
\mathcal{D}_{,\mathcal{N}}=\frac{3
   \mathcal{D}^2 \left[E^2 \left(3 \bar{Q}^2-1\right) \left(1-\bar{Q}\right)^2 \cos ^2(\theta )+3
   \bar{Q}^4\right]}{2  \bar{Q}\left(E^2-3 \bar{Q}^2\right)},
\end{equation}
The system \eqref{DynSysExpV2} has invariant submanifolds at  $\bar{Q}=\frac{1}{2} \left(\sqrt{3}-1\right)$, $\theta=k\pi$, $\theta=\pi/2 + k\pi$ with $k$ integer number. It is also singular in 
\begin{equation}
\bar{Q}=\frac{\cos (2 \theta )\pm2 \sqrt{3} \cos \theta +1}{\cos (2 \theta )-5}\qquad \bar{Q}=0\qquad \bar{Q}=1
\end{equation}
The system  \eqref{DynSysExpV2} admits  fixed points given by
\begin{equation}
\mathcal{A}_k\rightarrow\left[\frac{\pi}{2}+k\pi,\frac{1}{2}\left(1-\frac{1}{\sqrt{3}}\right)\right]
\end{equation}
where $k\in \mathds{Z}$ which are both saddles and associated with the solution 
\begin{equation}
a=a_0\left(t-t_0\right)^{2/3}.
\end{equation}
The above data analysis indicates the following values of the quantities appearing in the cosmological equations at present:  
\begin{equation}\label{ICV2}
\begin{split}
&\varepsilon \rightarrow 10^{-7},\quad H_0 \rightarrow  69.6 ,\quad  \rho_0 \rightarrow 4069.09 ,\quad \varphi _0 \rightarrow 0.005 ,\\ &V_0 \rightarrow -21116 ,\quad \lambda \rightarrow  1.5 ,\quad m \rightarrow 0.6
\end{split}
\end{equation}
In Fig. \ref{figPSV2} we have represented the compact phase space and the position representing the state of the Universe related to the data analysis of the previous sections. As in the case of the power law potential, the past attractor of the dynamics is a singular state on the invariant submanifold  and the Universe approaches the $\bar{Q}=1$ boundary after transiting close the fixed point $\mathcal{A}_1$. This is also confirmed by the time evolution of the variables in Fig. \ref{figDynVarV2}. The solution associated with the boundary $\bar{Q}=1$ is given by the equation
\begin{equation}
 \dot{\varphi}^2+V_0\frac{3}{2}\varphi^{2}\exp\left(-\lambda\varphi\right)=0
\end{equation}
whose solution can only be given parametrically. However, using the \eqref{14.1}, we can obtain
\begin{equation}
q=-\frac{\ddot{a}a}{\dot{a}^2}= -1-\frac{3\lambda\varphi_0}{2a^3}
\end{equation}
 which for growing $a$ shows that the solution to this equation corresponds to an accelerating cosmology, and indeed to an asymptotically de Sitter solution. This is also confirmed by  Fig. \ref{figQqV1} which plots the the behavior of the deceleration factor on the orbit selected by  the parameters \eqref{ICV2}.
Note also that different initial conditions  in this case lead to a final state with a different value of $\theta$.
 
\begin{figure}
\centering
\includegraphics[width=0.8\textwidth]{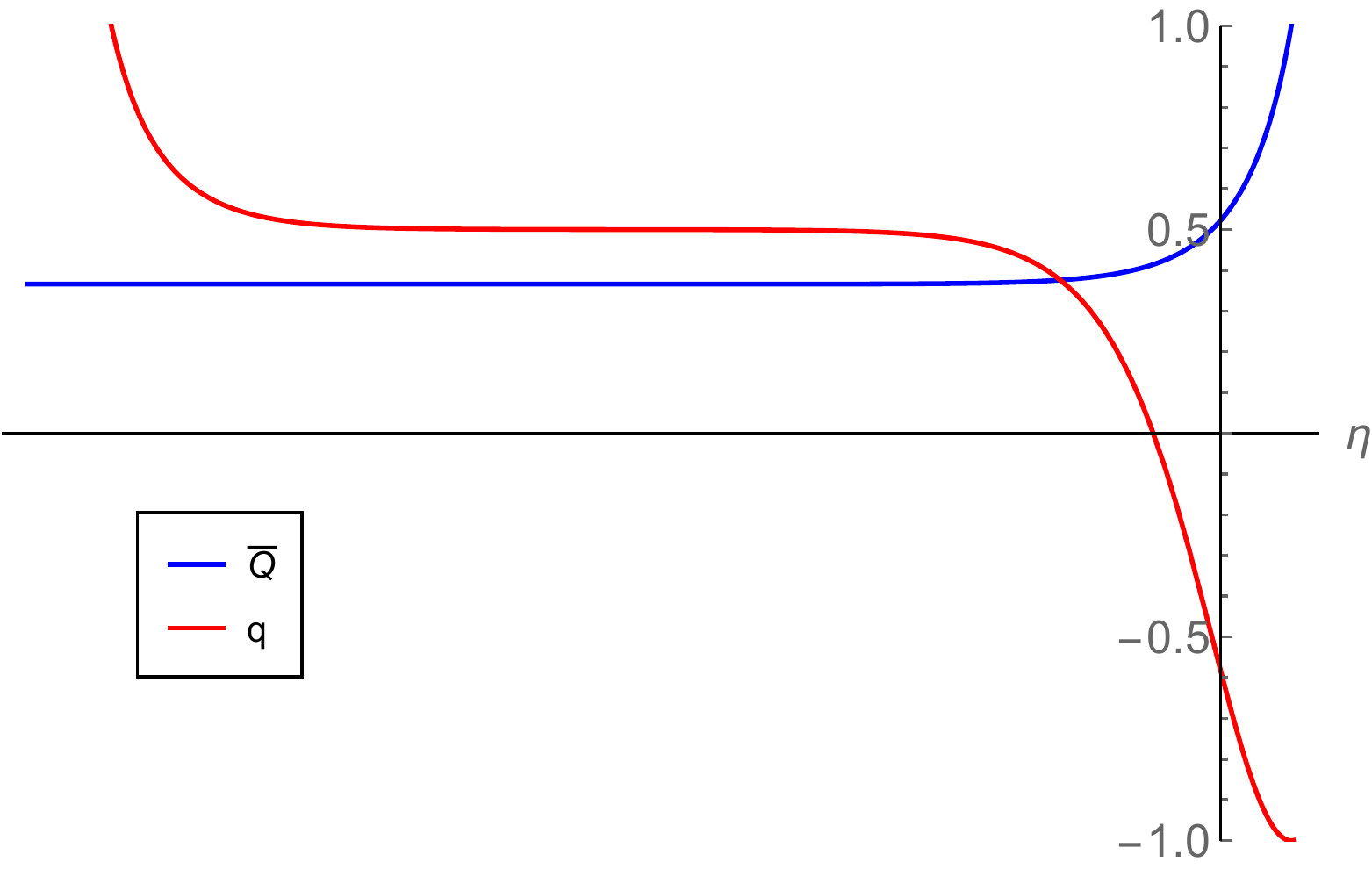}
\caption{Plot of the evolution of the variable $\bar{Q}$ and the deceleration factor $q$  along the orbits of the phase space for the potential $V(\varphi)=V_0 \exp\left(-\lambda\varphi\right)$ associated with the conditions \eqref{ICV2}. }
\label{figQqV2}
\end{figure}

\subsection{The case $V(\varphi)=V_0 (V_1 +\varphi)^\alpha$ } \label{PowerV3}
If one consider the exponential  potential $V(\varphi)=V_0 (V_1 +\varphi)^\gamma$, then
 \begin{equation}\label{DynPw2V3}
\begin{split}
  \bar{Q}_{,\mathcal{N}}&=-\frac{3 \bar{Q}^2 [2 \bar{Q} (\bar{Q}+1)-1] \left\{\bar{Q}^4
 V_1+(\bar{Q}-1)^4 \cos ^4(\theta ) [1+\gamma  (\cos (2 \theta )-1) ]\right\}}{2
   \left[(\bar{Q}-1)^2 \cos ^2(\theta )-3 \bar{Q}^2\right] \left[\bar{Q}^4
 V_1+(\bar{Q}-1)^4 \cos ^4(\theta )\right]},\\
 \theta_{,\mathcal{N}}&=\frac{3  \bar{Q}(\bar{Q}-1)
   \sin (\theta ) \cos (\theta ) \left\{\bar{Q}^4 V_1+(\bar{Q}-1)^2 \cos
   ^4(\theta ) \left[2 \gamma (2 \bar{Q}^2+2 \bar{Q}-1)+(\bar{Q}-1)^2\right]\right\}}{2 \left[(\bar{Q}-1)^2 \cos ^2(\theta )-3
   \bar{Q}^2\right] \left[\bar{Q}^4 V_1+(\bar{Q}-1)^4 \cos ^4(\theta )\right]},
\end{split}
\end{equation}
and 
\begin{equation}
\mathcal{D}_{,\mathcal{N}}=-\frac{3 \bar{Q} \mathcal{D}^2 \left\{3 \bar{Q}^6
   V_1+(\bar{Q}-1)^4 \cos ^4(\theta ) \left[3 \bar{Q}^2+2 \gamma  (2 \bar{Q}^2
   +2 \bar{Q}-1) \cos ^2(\theta )\right]\right\}}{2 (\bar{Q}-1) \left\{\left[(\bar{Q}-1)^2
   \cos ^2(\theta )-3 \bar{Q}^2\right] \left[\bar{Q}^4 V_1+(\bar{Q}-1)^4 \cos
   ^4(\theta )\right]\right\}}\,
\end{equation}
As expected the system above has much in common with the case  of Sec.~\ref{PowerV1}. Therefore we find the same invariant submanifolds, but additional singularities determined by the equation
\begin{equation}
V_1= -\frac{(\bar{Q}-1)^4 \cos ^4(\theta)}{\bar{Q}^4}
\end{equation}
The type of fixed points consist of a line and two isolated fixed points and they are the same of the ones in Sec.~\ref{PowerV1}, in agreement with the analysis of \cite{CVC} (see Table \ref{TableV3}). Also the stability of the isolated fixed points is the same. The points on the line, however, are unstable if
\begin{equation}
\gamma< 1/2 \quad -\frac{1}{\sqrt{2\gamma}}<\sin \theta< \frac{1}{\sqrt{2\gamma}}
\end{equation}

\begin{figure}
\centering
\includegraphics[width=0.8\textwidth]{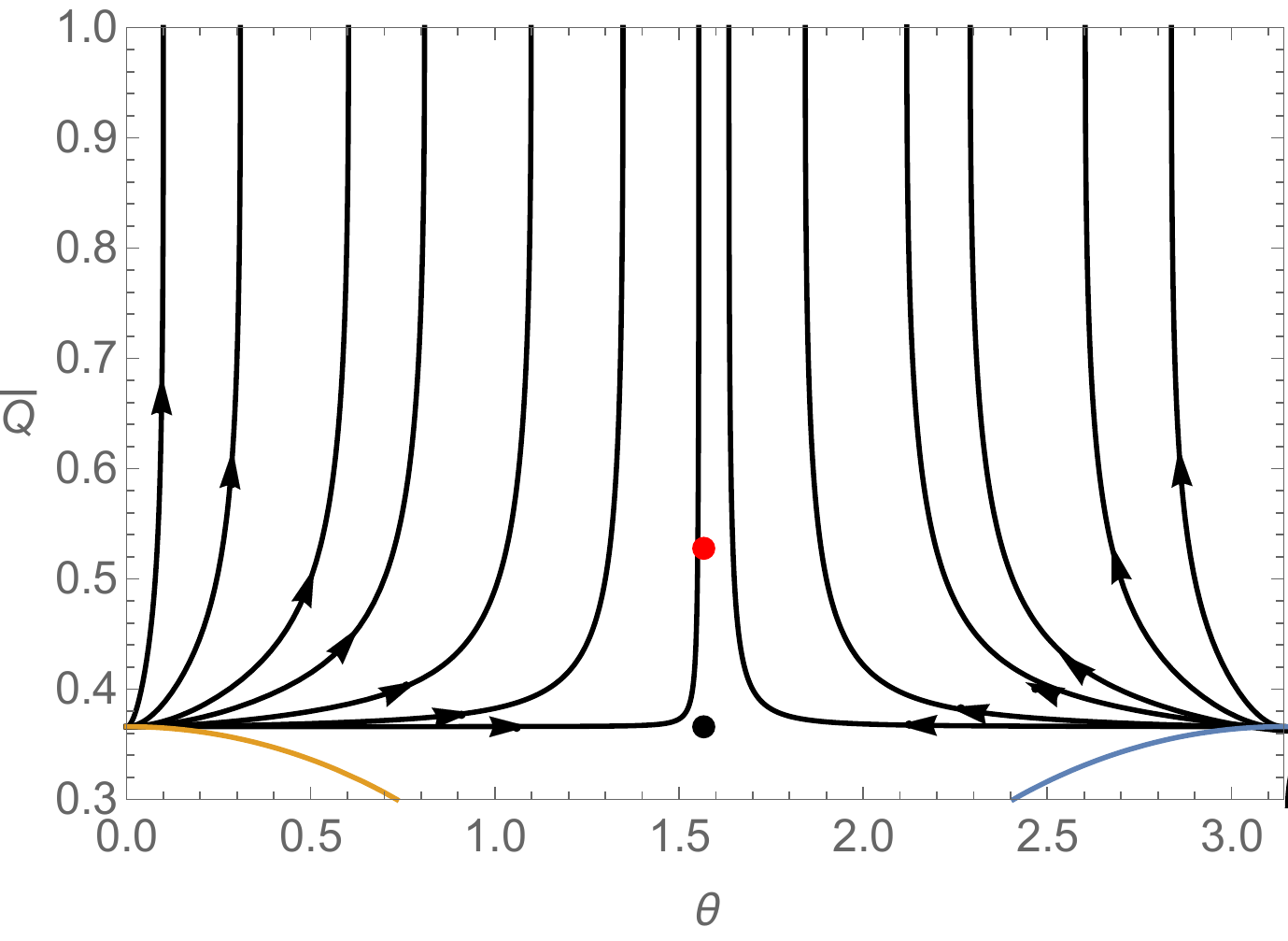}
\caption{Portion of the phase space of equations \eqref{DynPw2V3}. Here the black dot represents the fixed point $\mathcal{A}_1$ and the blue and blue lines the singularity of the dynamical system. The red dot represents the state of the universe as indicated from the observational analysis of Sections 4, 5 and 6.}
\label{figPSV3}
\end{figure}

\begin{figure}
\centering
\includegraphics[width=0.8\textwidth]{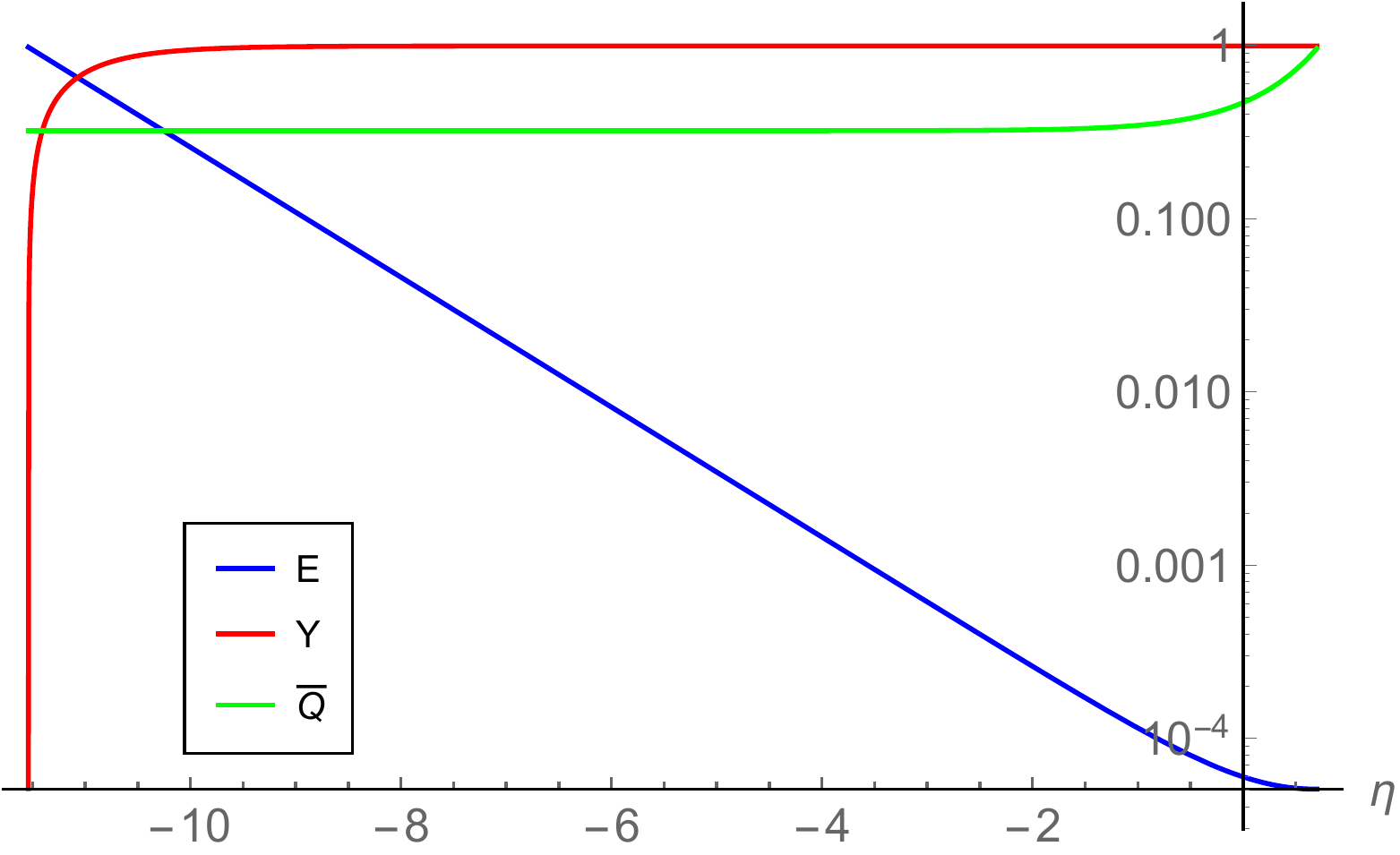}
\caption{Semilogarithmic plot of the evolution of the dynamical system variables along the orbits of the phasespace for the potential $V(\varphi)=V_0 (V_1 +\varphi)^\gamma$ associated with the conditions \eqref{ICV3}. }
\label{figDynVarV3}
\end{figure}

The above data analysis indicates the following values of the quantities appearing in the cosmological equations at present:  
\begin{equation}\label{ICV3}
\begin{split}
&\varepsilon \rightarrow 10^{-7},\quad H_0 \rightarrow  69.7 ,\quad  \rho_0 \rightarrow 4080.08 ,\quad \varphi _0 \rightarrow 0.005 ,\\ &V_0 \rightarrow -236 ,\quad \gamma \rightarrow  1.6 ,\quad m \rightarrow 0.55,\quad V_1 \rightarrow 16.5
\end{split}
\end{equation}
In Fig. \ref{figPSV3} we have represented the compact phase space and the position expressing the state of the Universe related to the data analysis of the previous sections.  As in the previous cases, the past attractor of the dynamics is a singular state reached coasting the invariant submanifold  $\bar{Q}=\frac{1}{2} \left(\sqrt{3}-1\right)$. 
The Universe approaches the $\bar{Q}=1$ boundary after transiting close the fixed point $\mathcal{A}_1$ as indicated by  the  evolution of the dynamical variables in time (see Fig. \ref{figDynVarV3}). Indeed, plotting the behavior of the deceleration factor (see Fig. \ref{figQqV1}) one can conclude that the observational data represent a cosmology which present cosmic acceleration. However, differently form the power law case of Sec.~\ref{PowerV1}, the cosmology in this case approaches a de Sitter evolution ($q=-1$) rather than an accelerated power law. As in Sec.~\ref{ExpV2}, we cannot give the solution corresponding the time asymptotic state  as the key equation
\begin{equation}
 \dot{\varphi}^2+V_0\frac{3}{2}\varphi^{2}(V_1 +\varphi)^\gamma=0
\end{equation}
does not have a solution that can be put in a useful analytical form, but, using eq. \eqref{14.1}, we can give the form of the decelerating factor
\begin{equation}
q=-\frac{\ddot{a}a}{\dot{a}^2}= -1+\frac{3 \gamma  \varphi_0^2}{a^6 V_1+\varphi_0^2-1}
\end{equation}
which approaches $-1$  for large $a$. As in the model of Sec.~\ref{ExpV2}, and in contrast with the model of Sec.~\ref{PowerV1}, different initial conditions lead to a final state with a different value of $\theta$.
\begin{table}[tbp] \centering
\caption{The fixed points and the solutions of the NMC model with matter and $V=V_0 (V_1 +\varphi)^\gamma$. Here $S$ stays for saddle, $A$ for attractor and $R$  for repeller.}
\begin{tabular}{cclllccc}
& & \\
\hline   Point &$[\theta,\bar{Q}]$ &  Scale Factor &  Stability  \\ \hline
& & \\
$\mathcal{A}_k$ &$\left[\frac{\pi}{2}+k\pi,\frac{1}{2}\left(1-\frac{1}{\sqrt{3}}\right)\right]$& $a=a_0\left(t-t_0\right)^{2/3}$&  S\\
& & \\
\multirow{2}{*}{$\mathcal{L}$}& \multirow{2}{*}{$\left(\theta_*,0\right)$}& \multirow{2}{*}{$a=a_0 t^{\frac{2 \cos\theta_*}{3 \left(\gamma -1\right)}}$} & R $\gamma< 1/2 \quad -\frac{1}{\sqrt{2\gamma}}<\sin \theta< \frac{1}{\sqrt{2\gamma}}$\\
& & & A otherwise\\
& & \\
 \hline
\end{tabular}\label{TableV3}
\end{table}
\begin{figure}
\centering
\includegraphics[width=0.8\textwidth]{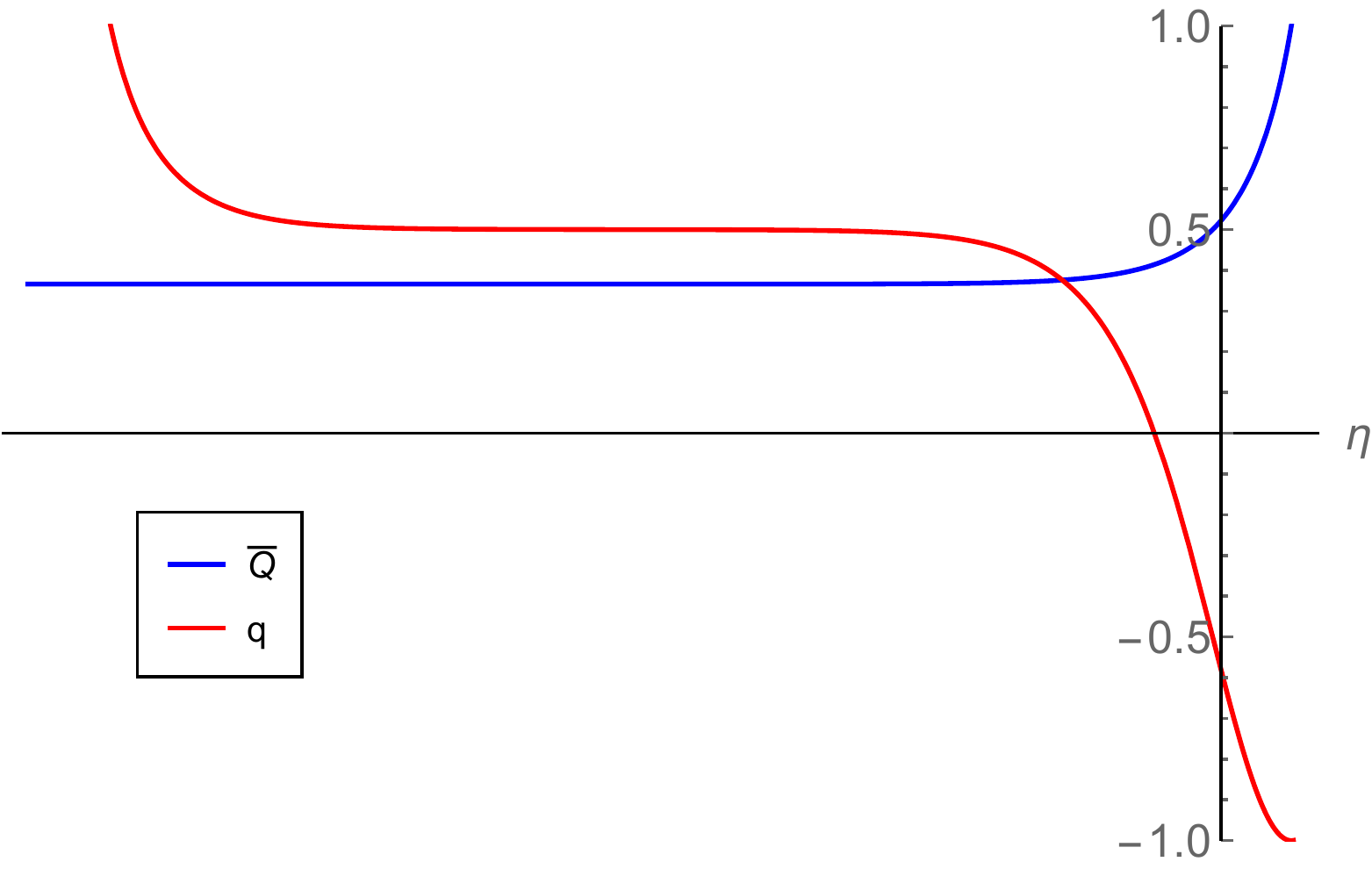}
\caption{Plot of the evolution of the variable $\bar{Q}$ and the deceleration factor $q$  along the orbit associated with the conditions \eqref{ICV3}. }
\label{figQqV3}
\end{figure}
\section{Conclusions}
In this paper we have analyzed some accelerating cosmological models in which dark energy is  modeled as a  non minimally coupled selfinteracting fermion condensate. Assuming standard matter to be non relativistic, we have investigated three different forms of potential $V(\varphi)$ that are important in cosmological models with scalar fields. 
 As a first example, we have considered the power law model $V(\varphi)=V_0 \varphi^\alpha$. At early times ($a\approx 0$ ) the \rf{18.1} implies that the scalar field will have high values and the potential will be negligible ($\alpha<0$) or dominant ($\alpha>0$). This situation is reversed at late time ($a\rightarrow\infty$ ). We have obtained a suitable parametrized form of the Hubble function written in terms of redshift, which,  with an appropriate choice of parameters, is comparable to the $\Lambda$CDM model within a wide range of redshifts and differs at very high redshift.

The second example we have considered has been the exponential potential $V(\varphi)=V_0 \exp(-\lambda \varphi)$ which is commonly used for scalar fields. At late time $\varphi\rightarrow0$ the exponential potential becomes actually a cosmological constant term: indeed, also in this case, the behavior of the function $H(z)$ mimics, for an appropriate choice of the parameters, the standard $\Lambda$CDM model.

The third example we have studied has been the potential $V=V_0 (\varphi^{2}+V_1)^\gamma$ which is a generalization of the power law potential. This potential combines characteristics of both the previous potentials as it behaves as a power law and generates an effective cosmological constant at late time, related to the value of the constant $V_1$.  
 After choosing appropriate parameterizations of these models, we have tested them with some different observational data. To this end, we have used the recently updated Supernovae Cosmology Project union 2.1, a sample of 193 GRB Hubble diagram, and a list of $28$ $H(z)$ measurements, compiled in \cite{farooqb}. Moreover, we have set Gaussian priors on the distance data from the BAO \cite{busca} and the Hubble costant $h$, in order to help break the degeneracies among the parameters of the different cosmological models.
For our statistical analysis, we have used the Markov Chain Monte Carlo Method (MCMC), running five parallel chains,  and using the Gelman-Rubin test to check the convergence. The histograms of the parameters from the merged chains have been then used to infer median values and confidence ranges. In Table \ref{tabcosmofit}, we have presented the results of our analysis for the main parameters concerning the three different models. 
The analysis we have performed indicates that a non--minimally coupled self-interacting fermion condensate is compatible with the datasets we have considered. 

 We have analyzed the three different models using the Akaike Information Criterion (AIC)  \cite{aic,aic2} and its indicator (\ref{aic}). We have found that  the model with the lower AIC is the the exponential potential one $V(\varphi)=V_0 \exp\left(-\lambda\varphi\right)$. We also used the same method to perform a comparison of the exponential potential model with the CPL parametrization for dark energy with an EOS given by (\ref{cpleos}). Interestingly, it turned out that the exponential model has the lower AIC indicating a weak evidence against the CPL model. This result should be confirmed using further data which will be available in the future.

 In addition, we have investigated the Euclid experimental possibility to constrain our model. We used simulated data assuming an Euclid\,-\,like survey of SNeIa \cite{Euclid}  (plotted in Fig.\,\ref{Euclidlikenz}) setting the exponential potential model as the fiducial model and considering the simulated SneIa HD  as cosmological probes. In Table \ref{SimSNexp} we have summarized the results of the simulation. These results indicate that the mock dataset is able to constraint much better our model.

Finally, using the values of the parameters deduced by the statistical analysis, we have used dynamical system techniques to infer the general behavior of the cosmology and in particular the behavior of the deceleration factor. As expected, it turns out that the three potentials lead to similar behaviors with differences in the final value of $q$ which is $-1$ for potentials that generate a cosmological constant term  and is higher than this value in the other cases. The behavior of the deceleration factor suggests that changes in the expansion rate during the matter dominate era might have repercussions in the large scale structure. Whether these differences are observables, however, cannot be deduced from our calculations and will require further work on the behavior of matter and condensate fluctuations. 

We conclude noticing that we have considered here standard matter in the form of dust and than the complete description of the cosmic history, in particular at earlier times, requires the inclusion of other forms of matter--energy, like radiation. This analysis is certainly possible with the dynamical system approach we proposed, but it requires a different treatment of the datasets, other than, probably additional sources of data. In addition, as we approach even earlier epochs the approximation of the semiclassical fermion which is the cornerstone of the theory we have considered starts to loose accuracy. This indicates that our approach to the analysis of this type of cosmology is useful but need to be extended with care in order to give meaningful results. Future works will consider this more complicated situation.

\section*{Acknowledgements}
EP acknowledges the support of INFN Sez. di Napoli  (Iniziativa Specifica QGSKY). SC was supported by an Investigador FCT Research contract through project IF/00250/2013 and acknowledges financial support provided under the European Union's H2020 ERC Consolidator grant ``Matter and strong-field gravity: New frontiers in Einstein's theory'' grant agreement No.~MaGRaTh646597, and under the H2020-MSCA-RISE-2015 grant No.~StronGrHEP-690904.
 The authors acknowledge the anonymous referee for her/his suggestions to improve this manunscript. 


\bibliographystyle{plain}

\end{document}